\documentclass[aps,prd,preprint, onecolumn, tightenlines, notitlepage, superscriptaddress, nofootinbib, preprintnumbers, floatfix]{revtex4-2}

\usepackage{amstext}
\usepackage{amssymb}
\usepackage{amsmath}
\usepackage{graphicx}
\graphicspath{{plots/}}
\usepackage{hyperref}
\usepackage{url}
\usepackage{color}
\usepackage{ulem}
\usepackage[utf8]{inputenc}
\pdfoutput=1
\usepackage[x11names]{xcolor}
\usepackage{textcomp}
\usepackage{subfigure}

\usepackage{epsfig,amsfonts,mathrsfs,amsmath,amssymb,graphicx,color,slashed,multirow}
\usepackage{amsmath,latexsym,amssymb,graphicx,slashed,hyperref,color,enumerate,url,cancel,gensymb}
\usepackage{textcomp}

\hypersetup{colorlinks,citecolor= nicegreen,linkcolor= nicegreen}
\definecolor{nicered}{rgb}{0.7,0.1,0.1}
\definecolor{nicegreen}{rgb}{0.1,0.5,0.1}

\def\beq{\begin{equation}}
\def\eeq{\end{equation}}
\def\beqn{\begin{eqnarray}}
\def\eeqn{\end{eqnarray}}

\definecolor{darkblue}{rgb}{0,0,80}

\AtBeginDocument{\hypersetup{citecolor=Green3,linkcolor=Blue1,urlcolor=Blue1}}

\begin{document}

\title{{\Large Large Extra Dimensions and neutrino experiments }}

`
\author{D.~V.~Forero}\email{dvanegas@udemedellin.edu.co}
\affiliation{Universidad de Medell\'{i}n, Carrera 87 N° 30 - 65 Medell\'{i}n, Colombia}

\author{C.~Giunti}
\email{carlo.giunti@to.infn.it}
\affiliation{Istituto Nazionale di Fisica Nucleare (INFN), Sezione di Torino, Via P. Giuria 1, I--10125 Torino, Italy}

\author{C.~A.~Ternes}\email{ternes@to.infn.it}
\affiliation{Istituto Nazionale di Fisica Nucleare (INFN), Sezione di Torino, Via P. Giuria 1, I--10125 Torino, Italy}

\author{O.~Tyagi}\email{oddhar89\_sps@jnu.ac.in}
\affiliation{Istituto Nazionale di Fisica Nucleare (INFN), Sezione di Torino, Via P. Giuria 1, I--10125 Torino, Italy}
\affiliation{School of Physical Sciences,  Jawaharlal Nehru University, 
      New Delhi 110067, India}

\begin{abstract}
The existence of Large Extra Dimensions can be probed in various neutrino experiments.
We analyze several neutrino data sets in a model with a dominant large extra dimension.
We show that the Gallium anomaly can be explained with neutrino oscillations induced by the large extra dimension,
but the region of parameter space which is preferred by the Gallium anomaly is in tension with the bounds from reactor rate data,
as well as the data of Daya Bay and MINOS.
We also present bounds obtained from the analysis of the KATRIN data.
We show, that current experiments can put strong bounds on the size
$R_{\text{ED}}$
of the extra dimension:
$R_{\text{ED}} < 0.20~\mu\text{m}$
and
$R_{\text{ED}} < 0.10~\mu\text{m}$
at 90\% C.L. for
normal and inverted ordering
of the standard neutrino masses, respectively.
\end{abstract}

\maketitle
\newpage
\tableofcontents

%%%%%%%%%%%%%%%%%%%%%%
%%%%%%%%%%%%%%%%%%%%%%
\section{Introduction}
\label{sec:intro}
%%%%%%%%%%%%%%%%%%%%%%
%%%%%%%%%%%%%%%%%%%%%%

The observation of neutrino oscillations
in many solar, atmospheric and long-baseline experiments
is an undisputed proof that neutrinos are massive and mixed particles.
The explanation of neutrino masses and mixing
requires the extension of the
Glashow-Weinberg-Salam
Standard Model (SM) in which neutrinos are massless
(see, e.g., the reviews in Refs.~\cite{Mohapatra:2006gs,Gonzalez-Garcia:2007dlo,Petcov:2013poa,King:2014nza}).
There are also other reasons to extend the SM.
A theoretical compelling reason is the explanation of the huge difference
between the electroweak and Planck scales,
which is called the ``hierarchy problem''.
A possible solution of the hierarchy problem is supplied by the existence of
Large Extra Dimensions (LED)~\cite{Arkani-Hamed:1998jmv,Arkani-Hamed:1998sfv},
which provides also an elegant explanation of the smallness of
neutrino masses~\cite{Arkani-Hamed:1998wuz,Dienes:1998sb,Dvali:1999cn,Mohapatra:2000wn,Barbieri:2000mg,Davoudiasl:2002fq}.
This is achieved with the introduction of
sterile right-handed neutrino fields which are singlets under the gauge symmetries of the SM and propagate in a space-time with $4+N_{\text{ED}}$ dimensions called the ``bulk'',
where $N_{\text{ED}}$ is the number of space-like extra dimensions.
The Yukawa couplings of
the right-handed neutrino fields
with the three standard left-handed active neutrino fields
$\nu_{e L}$,
$\nu_{\mu L}$,
$\nu_{\tau L}$
and the SM Higgs boson,
which are confined to the ordinary four-dimensional space-time
called the ``brane'',
generate Dirac neutrino masses
through the standard Higgs mechanism at the electroweak scale.
However,
the values of the Dirac neutrino masses are reduced with respect to the
electroweak scale by the suppression of the wave functions of the
right-handed neutrino fields on the brane
due to the volume of the large extra dimensions.

The existence of large extra dimensions
can be probed in neutrino experiments
through the effects of the Kaluza-Klein (KK)
excitations which describe the right-handed singlet neutrino fields
on the brane.
The phenomenology is similar to that of light sterile neutrinos
(see Ref.~\cite{Esmaili:2014esa}),
which induce observable short-baseline neutrino oscillations
and small perturbations to the neutrino oscillations
in solar, atmospheric and long-baseline experiments
that are well-described by standard three-neutrino mixing~\cite{deSalas:2020pgw,Esteban:2020cvm,Capozzi:2021fjo}.

It is common to consider a LED model with a dominant extra dimension~\cite{Davoudiasl:2002fq,Machado:2011jt,Machado:2011kt,Basto-Gonzalez:2012nel,Girardi:2014gna,Rodejohann:2014eka,Esmaili:2014esa,Berryman:2016szd,Carena:2017qhd,Stenico:2018jpl,Arguelles:2019xgp,DUNE:2020fgq,Basto-Gonzalez:2021aus,Arguelles:2022xxa}
whose neutrino phenomenology is described by only two parameters:
the neutrino mass scale and the radius of the dominant extra dimension.
In this paper we present the bounds on these parameters
obtained from the analysis of several neutrino oscillation experiments
and from the recent results of the KATRIN experiment
on the search for the effects of sub-eV neutrino masses
on the spectrum of electrons emitted in tritium decay.

We discuss the implications for the LED model
from the results of reactor short-baseline neutrino oscillation experiments
and the results of the Gallium source experiments
in the context of the reactor antineutrino anomaly~\cite{Mention:2011rk}
and the Gallium neutrino anomaly~\cite{Abdurashitov:2005tb,Laveder:2007zz,Giunti:2006bj}
(see, e.g., the reviews in Refs.~\cite{Gariazzo:2015rra,Giunti:2019aiy,Diaz:2019fwt,Boser:2019rta,Dasgupta:2021ies}).
In particular,
we take into account the recent results of the
BEST Gallium source experiment~\cite{Barinov:2021asz,Barinov:2022wfh}.

The plan of the paper is as follows:
in Section~\ref{sec:osc} we review the relevant aspects of neutrino mixing
in the LED model under consideration;
in Section~\ref{sec:data} we describe the method of analysis of the data
and the results of the separate analyses of the data of
short-baseline experiments (Subsection~\ref{sec:sbl}),
long-baseline experiments (Subsection~\ref{sec:lbl}),
and the KATRIN experiment (Subsection~\ref{sec:kat});
in Section~\ref{sec:comb} we present the results for the combined bounds on the LED parameters;
finally, in Section~\ref{sec:conc}
we present a summary of our results and the conclusions.

%%%%%%%%%%%%%%%%%%%%%%
%%%%%%%%%%%%%%%%%%%%%%
\section{Neutrino oscillations in presence of Large Extra Dimensions}
\label{sec:osc}
%%%%%%%%%%%%%%%%%%%%%%
%%%%%%%%%%%%%%%%%%%%%%

The possible existence of Large Extra Dimensions (LED)
was originally proposed as a solution of the hierarchy problem~\cite{Arkani-Hamed:1998jmv,Arkani-Hamed:1998sfv}.
In the LED model,
the ordinary four-dimensional space-time
is a brane embedded in a space-time with $4+N_{\text{ED}}$ dimensions,
having $N_{\text{ED}}$ large space-like extra dimensions.
The fields which are charged under the gauge symmetries of the Standard Model (SM)
are restricted to the four-dimensional brane,
whereas
the fields which are singlets under the SM gauge symmetries
propagate in the ($4+N_{\text{ED}}$)-dimensional bulk.
In particular,
right-handed sterile neutrino fields
are SM gauge singlets which propagate in the bulk.
The Yukawa couplings with the SM left-handed neutrinos
are suppressed by the LED volume,
leading to naturally small Dirac neutrino masses~\cite{Arkani-Hamed:1998wuz,Dienes:1998sb,Dvali:1999cn,Mohapatra:2000wn,Barbieri:2000mg,Davoudiasl:2002fq}.
As most phenomenological studies of neutrino physics in a LED model~\cite{Davoudiasl:2002fq,Machado:2011jt,Machado:2011kt,Basto-Gonzalez:2012nel,Girardi:2014gna,Rodejohann:2014eka,Esmaili:2014esa,Berryman:2016szd,Carena:2017qhd,Stenico:2018jpl,Arguelles:2019xgp,DUNE:2020fgq,Basto-Gonzalez:2021aus},
we consider a LED model with one of the extra dimensions which is compactified on
a circle with radius $R_{\text{ED}}$
which is much larger than the size of the other extra dimensions.
Therefore, we consider an effective five-dimensional space-time
and we assume that there are three five-dimensional
right-handed singlet fermion fields
associated with the three active left-handed flavor neutrino fields
$\nu_{\alpha L}$,
with $\alpha=e,\mu,\tau$.
Each of the five-dimensional
right-handed singlet fermion fields
can be decomposed
as an infinite tower of four-dimensional Kaluza-Klein (KK) fields.
After diagonalization of the mass matrix,
the mixing of the three active neutrinos is given by
\begin{equation}
\nu_{\alpha L}
=
\sum_{i=1}^3 U_{\alpha i}
\sum_{n=0}^{\infty} V_{in}
\nu_{i L}^{(n)}
,
\label{eq:mixing}
\end{equation}
where
$U$ is the ordinary $3 \times 3$ unitary neutrino mixing matrix
and each $\nu_{i L}^{(n)}$ is a neutrino field with mass
$m_{i}^{(n)} = \lambda_{i}^{(n)} / R_{\text{ED}}$,
where $\lambda_{i}^{(n)}$ are the solutions of the
transcendental equation
\begin{equation}
\lambda_{i}^{(n)}
-
\pi
\left( m_{i}^{\text{D}} R_{\text{ED}} \right)^2
\cot\!\left( \pi \lambda_{i}^{(n)} \right)
=
0
,
\label{eq:lambda}
\end{equation}
where
$m_{i}^{\text{D}}$
are the three eigenvalues of the Dirac neutrino mass matrix,
which are naturally much smaller than the electroweak scale
because of the LED volume suppression.
The components of the mixing matrix $V$ are given
by~\cite{Dienes:1998sb,Dvali:1999cn,Mohapatra:2000wn}
\begin{equation}
(V_{in})^2
=
\frac{2}{
1
+
\pi^2 \left( m_{i}^{\text{D}} R_{\text{ED}} \right)^2
+
( m_{i}^{(n)} / m_{i}^{\text{D}} )^2
}
.
\label{eq:V}
\end{equation}
The neutrino oscillation probability is given by
\begin{equation}
P_{\nu_\alpha\rightarrow \nu_\beta}
=
\left\vert
\sum_{i=1}^3
\sum_{n=0}^\infty
U_{\alpha i}^*
U_{\beta i}
V_{in}^2
\exp\!\left(
- i \frac{ (m_{i}^{(n)})^2 L }{ 2 E }
\right)
\right\vert^2
,
\label{eq:oscprob}
\end{equation}
where $E$ is the neutrino energy
and $L$ is the source-detector distance.

The transcendental equation~\eqref{eq:lambda}
has an infinite number of solutions
$\lambda_{i}^{(n)}$
for $n=0,1,\ldots,\infty$
in the intervals
$[n,n+1/2]$.
To get a feeling of the behavior of these solutions,
one can solve the transcendental equation~\eqref{eq:lambda}
analytically for
$m_{i}^{\text{D}} R_{\text{ED}} \ll 1$~\cite{Davoudiasl:2002fq}
and find the leading expressions
\begin{align}
\null & \null
m_{i}^{(0)}
=
m_{i}^{\text{D}}
\left[
1
-
\frac{\pi^2}{6}
\,
\left( m_{i}^{\text{D}}  R_{\text{ED}} \right)^2
+
\ldots
\right]
,
\label{eq:mi0}
\\
\null & \null
m_{i}^{(k)}
=
\frac{k}{R_{\text{ED}}}
\left[
1
+
\frac{ \left( m_{i}^{\text{D}} R_{\text{ED}} \right)^2 }{ k^2 }
+
\ldots
\right]
\quad
\text{for}
\quad
k > 0
,
\label{eq:mik}
\\
\null & \null
V_{i0}
=
1
-
\frac{\pi^2}{6}
\,
\left( m_{i}^{\text{D}}  R_{\text{ED}} \right)^2
+
\ldots
,
\label{eq:Vi0}
\\
\null & \null
V_{ik}
=
\sqrt{2}
\,
\frac{ m_{i}^{\text{D}} R_{\text{ED}} }{ k }
\left[
1
-
\frac{3}{2}
\,
\frac{ \left( m_{i}^{\text{D}} R_{\text{ED}} \right)^2 }{ k^2 }
+
\ldots
\right]
\quad
\text{for}
\quad
k > 0
,
\label{eq:Vik}
\end{align}
Therefore, for $k>0$
the masses
$m_{i}^{(k)}$
increase with $k$
and
the mixing $V_{ik}$
decreases with $k$.

\begin{figure}[t!]
\centering
\subfigure[]{\label{fig:NO-m0-mass}
\includegraphics*[width=0.48\textwidth]{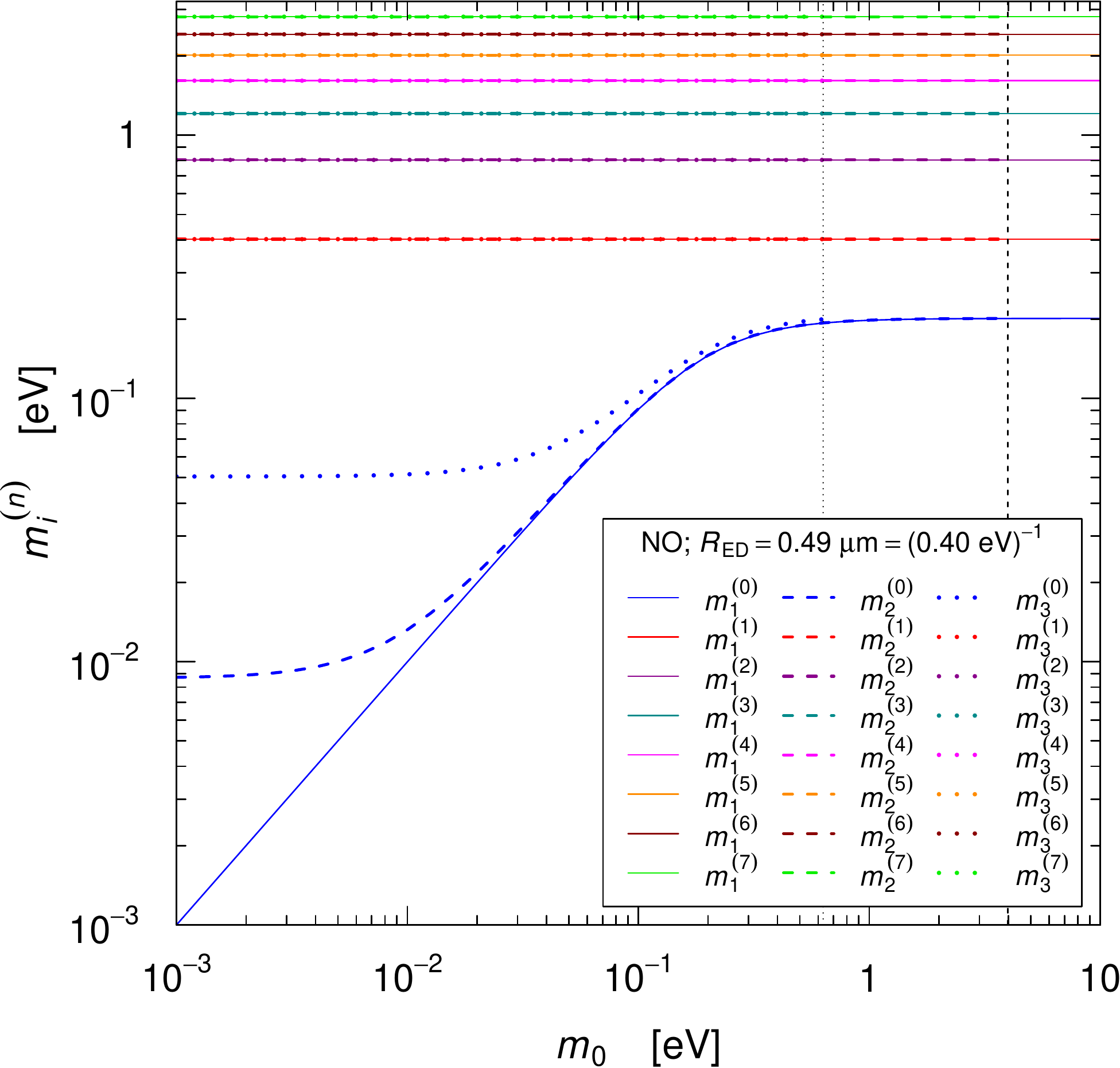}
}
\subfigure[]{\label{fig:IO-m0-mass}
\includegraphics*[width=0.48\textwidth]{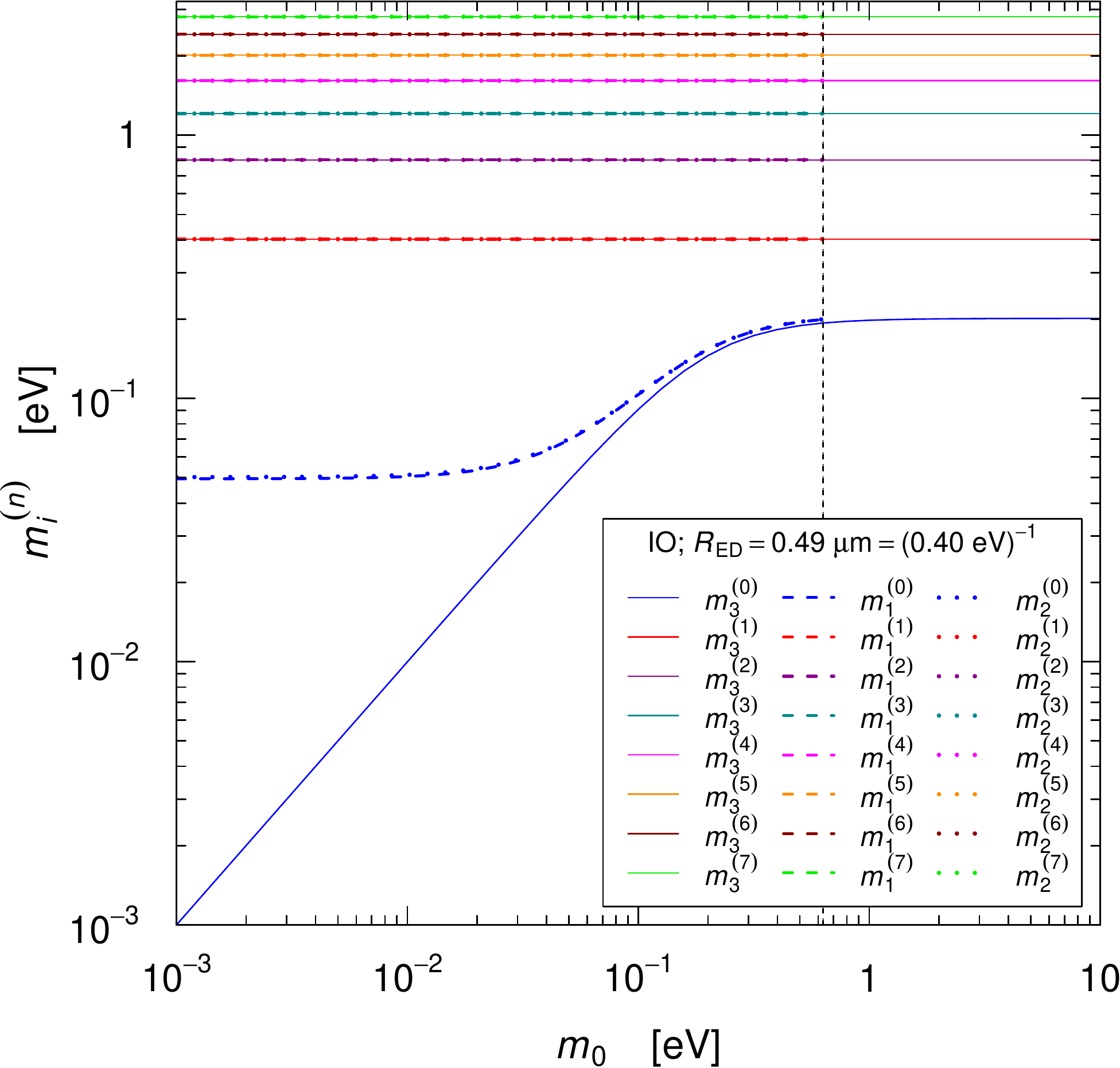}
}
\caption{\label{fig:m0-mass}
The neutrino masses $m_{i}^{(n)}$
for
$n=0,1,\ldots,7$,
as functions of $m_{0}$
for
$R_{\text{ED}}=0.49~\mu{m}$
in the
\subref{fig:NO-m0-mass} Normal Ordering ($m_{0} \equiv m_{1}^{\text{D}}$)
and
\subref{fig:IO-m0-mass} Inverted Ordering ($m_{0} \equiv m_{3}^{\text{D}}$).
The dashed and dotted vertical lines indicate the largest value of $m_{0}$
determined by the inequality~\eqref{eq:physical}
for $r=2$ and $r=3$ in NO (where $s=1$)
and
for $r=1$ and $r=2$ in IO (where $s=3$).
Note that in IO all the dashed and dotted lines are almost superimposed,
because the differences are determined by the small solar mass-squared difference.
} 
\end{figure}

Since the standard three-neutrino mixing
describes well the oscillations
observed in solar, atmospheric and long-baseline neutrino experiments,
the LED model must be considered as a perturbation of
three-neutrino mixing,
which corresponds to $V_{in}=\delta_{n0}$ for $i=1,2,3$.
Hence,
we require that the zero-mode masses $m_{i}^{(0)}$ generate the standard
mass-squared differences~\cite{Basto-Gonzalez:2012nel}:
\begin{equation}
\Delta{m}^2_{kj}
=
(m_{k}^{(0)})^2 - (m_{j}^{(0)})^2
.
\label{eq:dm2}
\end{equation}
We consider ~\cite{deSalas:2020pgw}:
\begin{equation}
\begin{split}
\Delta{m}^2_{21} &= 7.5\times10^{-5}~\text{eV}^2
,
\\
(\Delta{m}^2_{31})_{\text{NO}} &= 2.55\times10^{-3}~\text{eV}^2
,
\\
(\Delta{m}^2_{31})_{\text{IO}} &= -2.45\times10^{-3}~\text{eV}^2
,
\end{split}
\label{eq:standard}
\end{equation}
where NO and IO indicate, respectively,
Normal Ordering
and
Inverted Ordering.
In this way, the two independent squared-mass mass-squared differences
$\Delta{m}^2_{21}$ and $\Delta{m}^2_{31}$
allow us to fix two of the independent Dirac mass parameters
$m_{i}^{\text{D}}$
of the LED model.
It is convenient to choose as the remaining free mass parameter,
denoted by $m_{0}$,
the lightest Dirac mass,
which depends on the ordering:
$m_{0} = m_{1}^{\text{D}}$ ($m_{0} = m_{3}^{\text{D}}$) in the normal (inverted) neutrino mass ordering.
With this method,
the LED model depends on two parameters:
$R_{\text{ED}}$ and $m_0$.
For fixed values of $R_{\text{ED}}$ and $m_0$,
the determination of the masses and mixing is done as follows.
First we determine the lightest zero-mode mass
$m_{1}^{(0)}$ ($m_{3}^{(0)}$)
by solving the transcendental equation in Eq.~\eqref{eq:lambda}
with
$ m_{1}^{\text{D}} = m_{0} $ ($ m_{3}^{\text{D}} = m_{0} $)
for NO (IO).
Next,
we determine the other two zero-mode masses
using Eq.~\eqref{eq:dm2}
and the values of the mass-squared differences in Eq.~\eqref{eq:standard}.
Then,
the transcendental equation~\eqref{eq:lambda}
allows us to determine the corresponding $m_{i}^{\text{D}}$'s.
In this way,
the values of all the three $m_{i}^{\text{D}}$'s are established
and
we can calculate the masses and mixings for all values of $n>1$
using Eqs.~\eqref{eq:lambda} and~\eqref{eq:V}.

The constraints~\eqref{eq:dm2} restrict the physical region of the
LED parameters $m_{0}$ and $R_{\text{ED}}$,
that must allow the $\lambda_{i}^{(0)}$'s to be smaller than 1/2
in order to have a solution of the transcendental equation~\eqref{eq:lambda}.
Denoting with $r$ and $s$ the indices of the largest and smallest zero-mode mass
($r=3$ and $s=1$ in NO; $r=2$ and $s=3$ in IO)
the physical region is determined by the inequality
\begin{equation}
R_{\text{ED}}^2 \Delta{m}^2_{rs}
+
(\lambda_{s}^{(0)})^2
=
(\lambda_{r}^{(0)})^2
<
1/4
.
\label{eq:physical}
\end{equation}
Note that there is a bound even for $m_{0} R_{\text{ED}} \ll 1$,
which can be found using the approximation in Eq.~\eqref{eq:mi0}:
\begin{equation}
R_{\text{ED}}
<
1 / (2 \sqrt{\Delta{m}^2_{rs}})
.
\label{eq:Rbound}
\end{equation}
Since the largest mass-squared difference is the atmospheric one,
which is about $2.5\times10^{-3}~\text{eV}^2$
in both NO and IO,
we have the physical upper bound for the radius of the extra dimension
\begin{equation}
R_{\text{ED}}
\lesssim
2~\mu\text{m}
.
\label{eq:Rlimit}
\end{equation}

\begin{figure}[t!]
\centering
\subfigure[]{\label{fig:NO-m0-vmix}
\includegraphics*[width=0.48\textwidth]{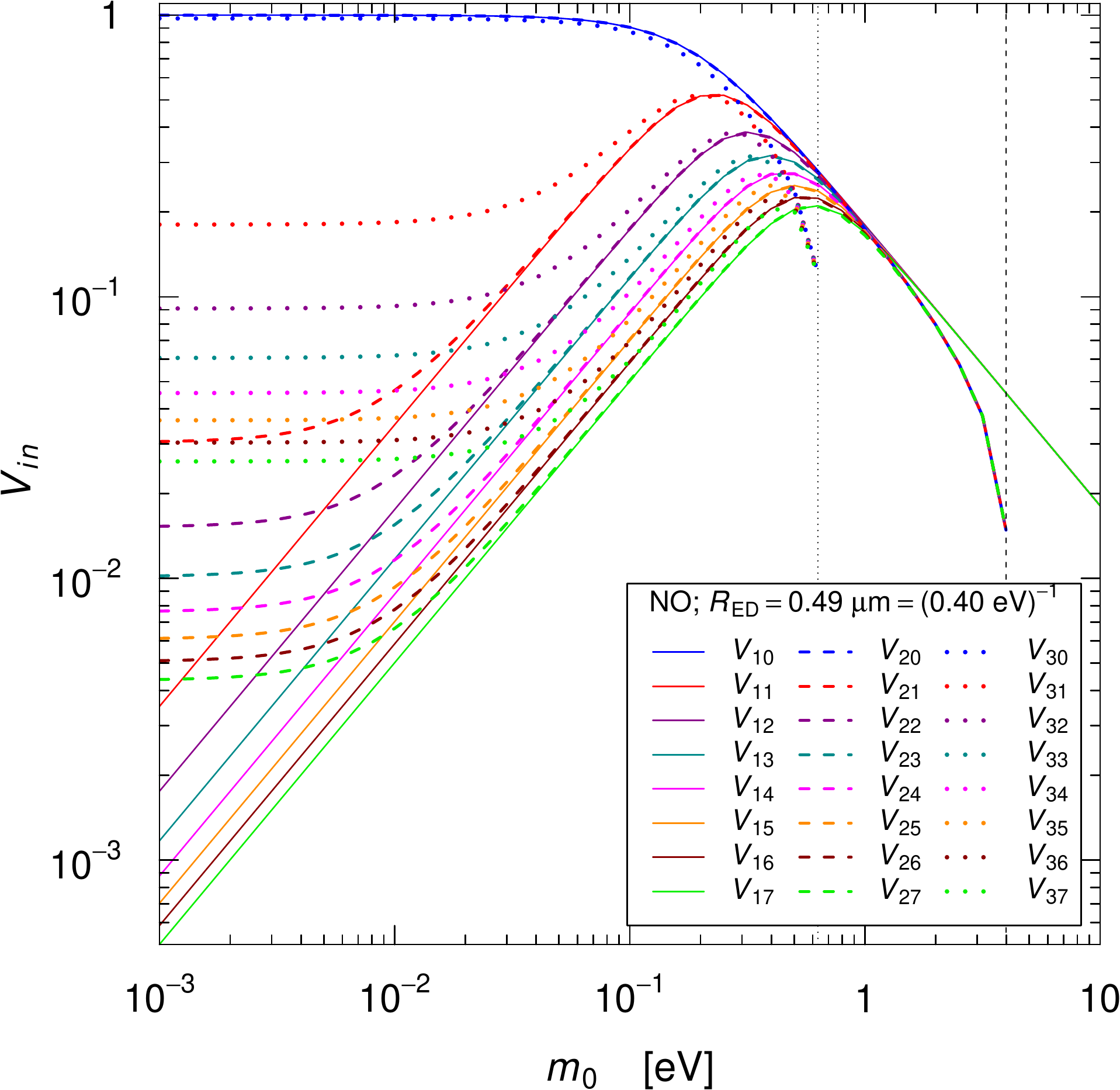}
}
\subfigure[]{\label{fig:IO-m0-vmix}
\includegraphics*[width=0.48\textwidth]{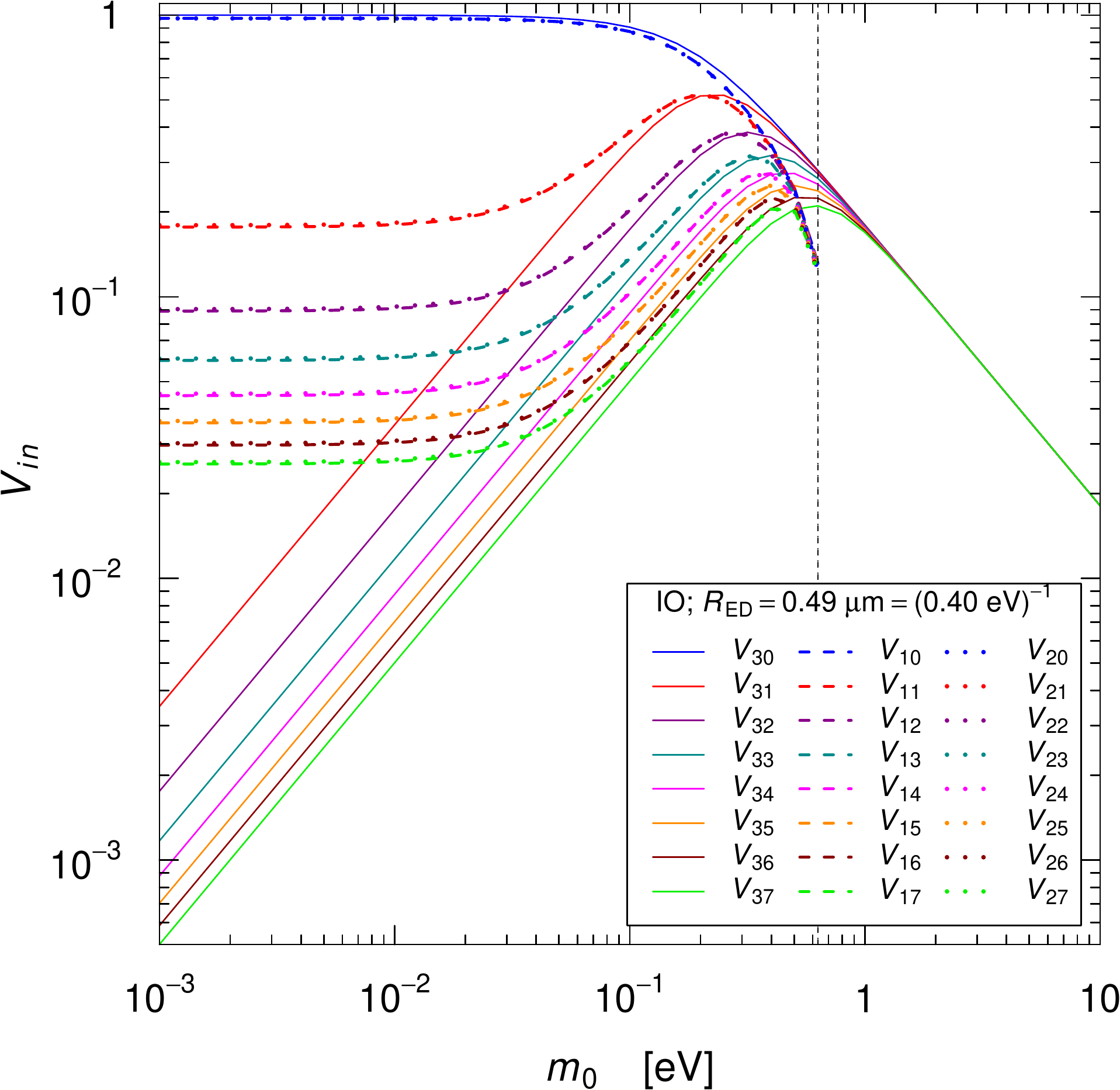}
}
\caption{\label{fig:m0-vmix}
The mixing elements $V_{in}$
for
$n=0,1,\ldots,7$,
as functions of $m_{0}$
for
$R_{\text{ED}}=0.49~\mu{m}$
in the
\subref{fig:NO-m0-vmix} Normal Ordering ($m_{0} \equiv m_{1}^{\text{D}}$)
and
\subref{fig:IO-m0-vmix} Inverted Ordering ($m_{0} \equiv m_{3}^{\text{D}}$).
The dashed and dotted vertical lines indicate the largest value of $m_{0}$
determined by the inequality~\eqref{eq:physical}
for $r=2$ and $r=3$ in NO (where $s=1$)
and
for $r=1$ and $r=2$ in IO (where $s=3$).
Note that in IO all the dashed and dotted lines are almost superimposed,
because the differences are determined by the small solar mass-squared difference.
} 
\end{figure}

The behavior of the masses
is illustrated in Fig.~\ref{fig:m0-mass}
for the zero-mode and the first seven KK modes.
The masses are plotted as functions of
$m_{0}$
for
$R_{\text{ED}}=0.49~\mu{m}=(0.40~\text{eV})^{-1}$.
One can see that for small values of $m_{0}$
the smallest zero-mode mass is approximately equal to $m_{0}$,
in agreement with Eq.~\eqref{eq:mi0}.
There is a deviation from this equality for
$m_{0} \gtrsim 0.1~\text{eV}$
and the smallest zero-mode mass tends to
$1/(2R_{\text{ED}}) = 0.20~\text{eV}$ for $m_{0} \gtrsim 1~\text{eV}$.
The dashed and dotted vertical lines indicate the largest value of $m_{0}$
determined by the inequality~\eqref{eq:physical}
for the other two zero-mode masses.
The physical upper bound on $m_{0}$ is determined by the
dashed lines,
that correspond to $m_{0} \lesssim 0.63~\text{eV}$
in both NO and IO.

Figure~\ref{fig:m0-vmix}
illustrates the behaviour of the mixing elements
$V_{in}$
corresponding to the masses in Fig.~\ref{fig:m0-mass}.
One can see that for small values of $m_{0}$
the mixing elements
$V_{i0}$
are dominant and the mixing elements
$V_{ik}$
decrease with $k$, in agreement with Eq.~\eqref{eq:Vik}.
Therefore,
for $m_{0} \ll R_{\text{ED}}^{-1}$
the phenomenology of the LED model is determined by a few low KK modes.
On the other hand,
for $m_{0} \gtrsim R_{\text{ED}}^{-1}$
it is necessary to take into account
a sufficient number of high KK modes. We have verified that for the oscillation experiments considered in this paper adding more than 5 modes does not effect our results. For the analyses of oscillation data we therefore use 5 modes. In the case of KATRIN one needs to calculate the number of relevant modes dependent on the values of $m_0$ and $R_{\text{ED}}$, as will be explained below in Sec.~\ref{sec:kat}.

%%%%%%%%%%%%%%%%%%%%%%
%%%%%%%%%%%%%%%%%%%%%%
\section{Data analysis}
\label{sec:data}
%%%%%%%%%%%%%%%%%%%%%%
%%%%%%%%%%%%%%%%%%%%%%
In this section we present the results of our analyses. Subsection~\ref{sec:sbl} discusses the analysis of reactor rate and Gallium experiments, Subsection~\ref{sec:lbl} is dedicated to MINOS/MINOS+ and Daya Bay, and in Subsection~\ref{sec:kat} we outline the analysis procedure of KATRIN data. In Section~\ref{sec:comb} we will discuss the combined bound from all experiments. 

When performing the analyses we take into account possible correlations between LED and standard oscillation parameters by marginalizing over them. In particular, marginalizing over $\theta_{13}$ and $\theta_{23}$ affects the bounds obtained in the analyses of Daya Bay and MINOS/MINOS+ data, respectively. 
Since current data only slightly prefer normal over inverted neutrino mass ordering~\cite{Gariazzo:2022ahe} of the three mostly active states, we perform all of our analyses for both orderings.

For electron neutrino experiments
($\nu_e$ disappearance and $\beta$-decay experiments),
the effects of the LED KK modes are quite different
in the two orderings for small values of $m_0$.
As one can see from Fig.~\ref{fig:NO-m0-mass},
in the NO case
$m_1^{(0)},m_2^{(0)} \ll m_3^{(0)}$
for $m_0 \ll 0.1~\text{eV}$.
Hence,
from Eq.~\eqref{eq:mi0} it follows that
$m_1^D, m_2^D \ll m_3^D$
for
$m_i^D R_{\text{ED}} \ll 1$.
Then,
from Eq.~\eqref{eq:Vik}
we obtain
$V_{1k},V_{2k} \ll V_{3k}$.
Therefore in the NO case the dominant LED effects
are given by the terms with $i=3$
in the neutrino oscillation probability \eqref{eq:oscprob}
for $\nu_e$ disappearance
and in $\beta$-decay spectra as that of the KATRIN experiment
that we consider in Section~\ref{sec:kat}.
Since these terms are suppressed by the smallness of $|U_{e3}|^2$,
the LED effects are small.
On the other hand,
in the IO case
$m_3^{(0)} \ll m_1^{(0)},m_2^{(0)}$
for $m_0 \ll 0.1~\text{eV}$,
as one can see from Fig.~\ref{fig:IO-m0-mass},
which implies that
$m_3^D \ll m_1^D, m_2^D$
and
$V_{3k} \ll V_{1k},V_{2k}$
for
$m_i^D R_{\text{ED}} \ll 1$.
In this case,
the dominant LED contributions appear in the terms with $i=1$ and $i=2$, which are not suppressed.
Therefore, 
$\nu_e$ disappearance and $\beta$-decay experiments
give stronger constraints on the LED parameters
for inverted ordering than for normal ordering
(see, e.g., Refs~\cite{Basto-Gonzalez:2012nel,Basto-Gonzalez:2021aus}).
Since this occurs for $m_0 \ll 0.1~\text{eV}$,
the result is a significant decrease of
the upper bounds for $R_{\text{ED}}$
for small values of $m_0$
in the case of inverted ordering with respect to normal ordering.
Instead,
in the case of $\nu_\mu$ disappearance all the relevant $|U_{\mu i}|^2$ are large
and there is no big difference between normal and inverted ordering.

%%%%%%%%%%%%%%%%%%%%%%
\subsection{LED at short baseline experiments}
\label{sec:sbl}
%%%%%%%%%%%%%%%%%%%%%%

In this section we present the results of our analyses of the data of the reactor rate experiments and the Gallium source experiments.
It has been shown in Ref.~\cite{Machado:2011kt} that the Gallium and reactor anomalies can be resolved if LED induced neutrino oscillations are present in nature. In this section we perform an updated analysis using the most up-to-date reactor rate and Gallium data. 

The importance of reactor fluxes and its impact on the statistical significance of the anomaly has been recently discussed in Refs.~\cite{Giunti:2021kab,Berryman:2020agd}. Here we repeat the reactor rate analysis as in Ref.~\cite{Giunti:2021kab}, but with the 3+1 oscillation probability replaced by its LED counterpart. The flux models that are considered in this work are the Huber-Mueller model~\cite{Mueller:2011nm,Huber:2011wv} (HM), the model by Estienne, Fallot \textit{et al}~\cite{Estienne:2019ujo} (EF), the model by Hayen \textit{et al}~\cite{Hayen:2019eop} (HKSS) and the recent model from the Kurchatov institute~\cite{Kopeikin:2021ugh} (KI). For a detailed description of the models and for the details of the statistical analysis the interested reader is referred to Ref.~\cite{Giunti:2021kab}.

\begin{figure}
  \centering
  \includegraphics[width=0.49\textwidth]{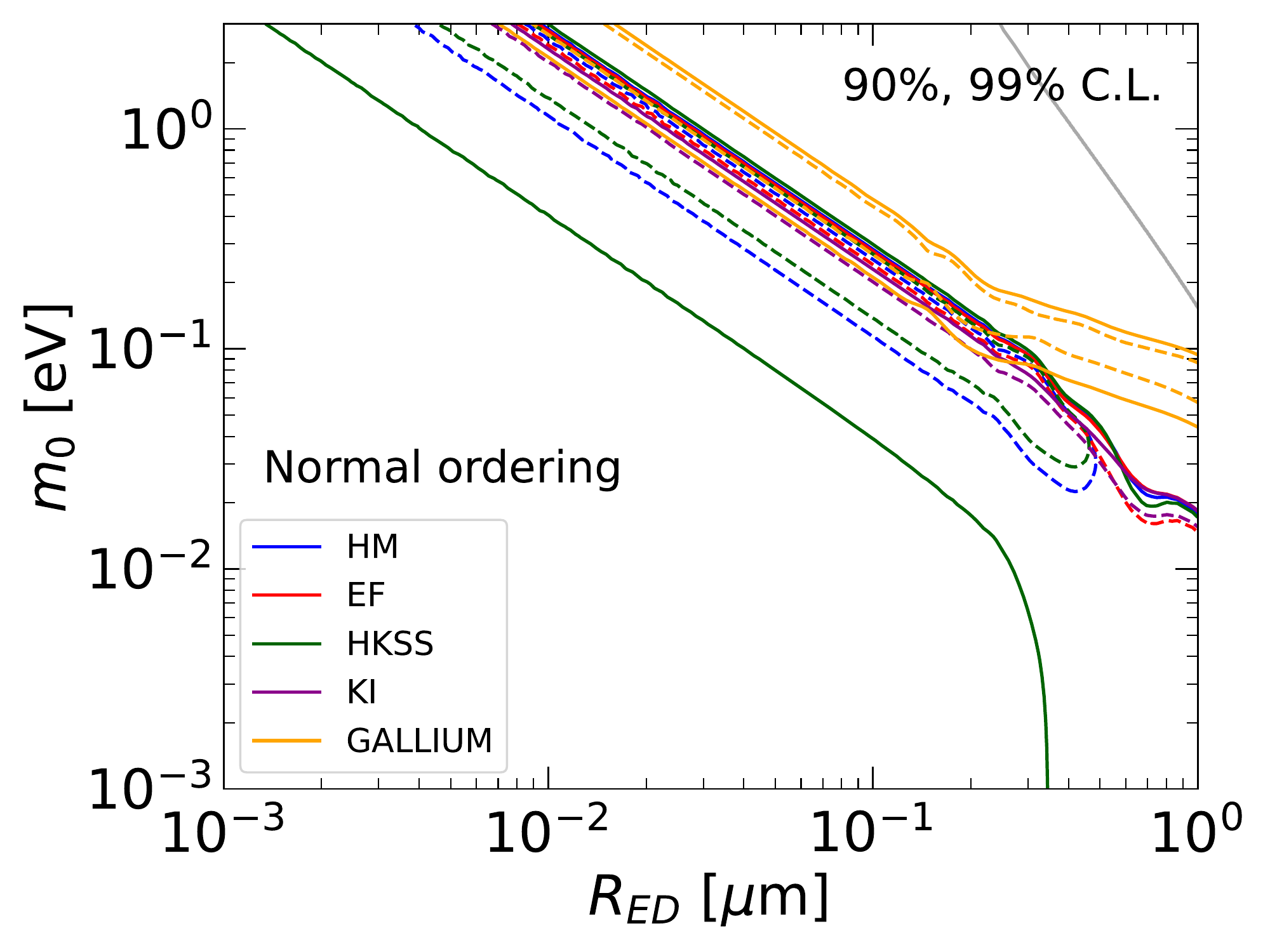}
    \includegraphics[width=0.49\textwidth]{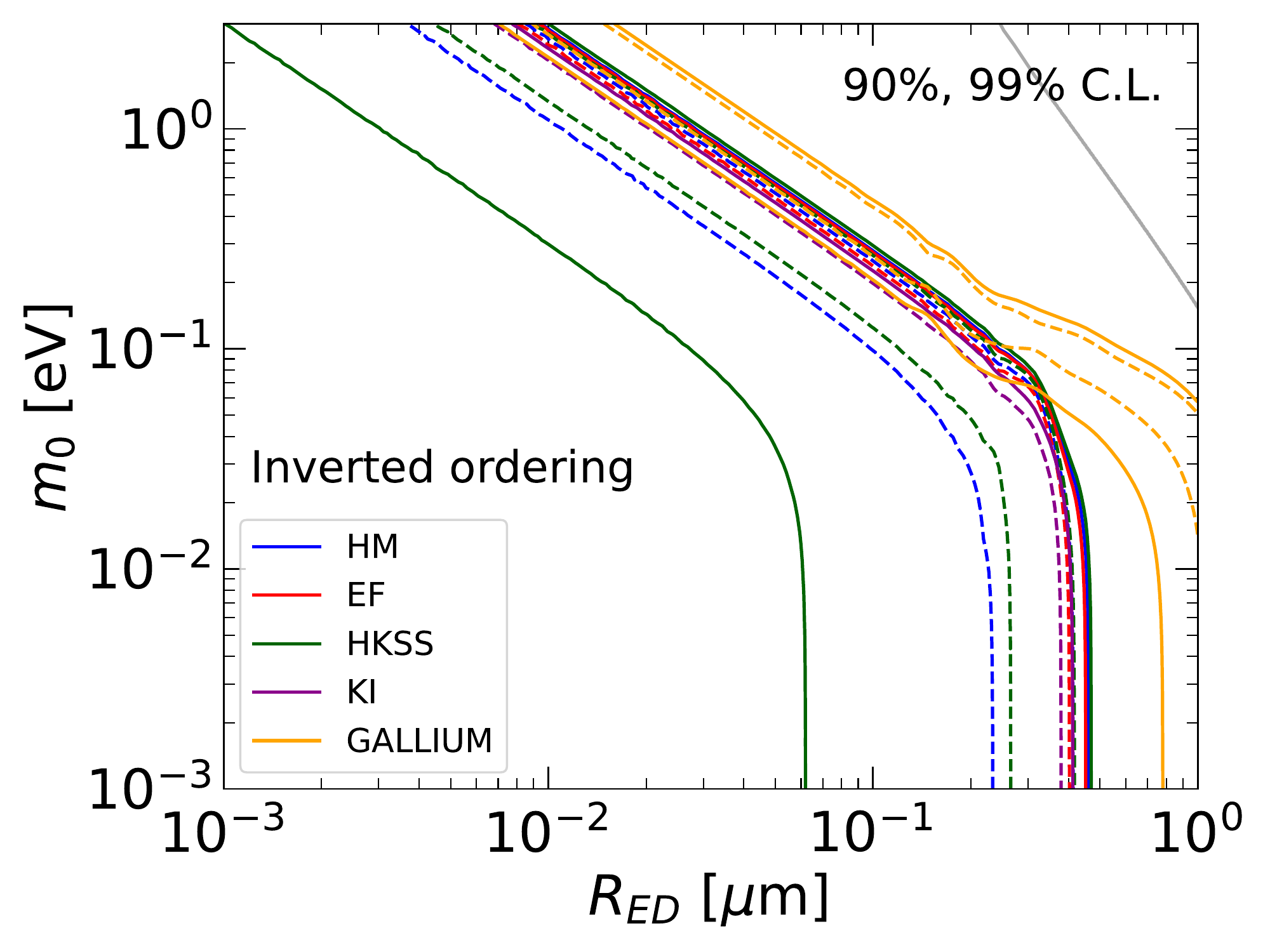}
    \caption{Results obtained from the analysis of reactor rate data for several models of the reactor antineutrino flux and the analysis of the Gallium data for normal (left) and inverted (right) neutrino mass ordering. The gray lines in the upper right corner mark the limit of the physically allowed region of parameter space.} 
  \label{fig:sbl}
\end{figure}

The results for the four flux models are shown in Fig.~\ref{fig:sbl} for normal (left panel) and inverted (right panel) neutrino mass ordering. The results are similar to those in Ref.~\cite{Giunti:2021kab}. For the KI and EF models no anomaly is found and we can only set bounds on the LED parameter space. In the case of the HM model we find an elongated region at 90\% C.L., but no closed regions at 99\% C.L. In the case of the HKSS model the anomaly is the strongest and then also the 99\% C.L. contour is closed. 

The GALLEX~\cite{GALLEX:1994rym,GALLEX:1997lja,Kaether:2010ag} and SAGE~\cite{Abdurashitov:1996dp,SAGE:1998fvr,Abdurashitov:2005tb,SAGE:2009eeu} experiments were constructed to detect solar neutrinos. They have been tested by placing intense artificial $^{51}$Cr and $^{37}$Ar radioactive sources inside of the detectors. The resulting ratios of observed to expected events are significantly smaller than unity. This deficit became known as the Gallium anomaly~\cite{Abdurashitov:2005tb,Laveder:2007zz,Giunti:2006bj}. The results of GALLEX and SAGE have been recently confirmed by the BEST collaboration~\cite{Barinov:2021asz,Barinov:2022wfh},  pushing the combined significance of the Gallium anomaly to the $5\sigma$ level. We analyze the data of the Gallium experiments following Ref.~\cite{Acero:2007su}, but including also the BEST data. We also consider a 3\% uncertainty on the Gallium cross section. We use only the cross section model of Bahcall~\cite{Bahcall:1997eg}, since other models produce a similar Gallium anomaly~\cite{Berryman:2021yan}.
In Fig.~\ref{fig:sbl} we show the preferred region obtained from the combined analysis of the data of the Gallium experiments (orange lines). This region updates the region obtained in Ref.~\cite{Machado:2011kt} by including also the recent results from the BEST experiment~\cite{Barinov:2021asz}. 
One sees, that the results of the reactor rate analysis, for all the considered flux models, and the result from the Gallium analysis are in tension. The same feature has already been observed in the context of 3+1 oscillations, see Ref.~\cite{Giunti:2021kab}.

%%%%%%%%%%%%%%%%%%%%%%
\subsection{LED at long baseline experiments}
\label{sec:lbl}
%%%%%%%%%%%%%%%%%%%%%%

The Main Injector Neutrino Oscillation Search (MINOS) was an accelerator-based neutrino oscillation experiment. 
Unlike the former experiments considered in this paper, MINOS uses a beam of muon neutrinos, instead of electron neutrinos.
The neutrinos were produced at the NuMI beam facility at Fermilab and detected at the near and far detectors of the experiment, located at 1.04~km and 735~km, respectively. During the MINOS data taking period, the neutrino beam peaked at an energy of 3~GeV and was later tuned to cover larger energies peaking at 7~GeV for the upgraded version of the experiment, MINOS+.
Traces of Large Extra Dimensions in the data have been sought for by the MINOS collaboration~\cite{MINOS:2016vvv}.
We update these results, considering the data corresponding to an exposure of $10.56\times10^{20}$ POT in MINOS (mostly in neutrino mode, only $3.36\times10^{20}$ POT were gathered in antineutrino mode) and $5.80\times10^{20}$ POT in MINOS+ (in neutrino mode). The data are the same as those used for the search of light sterile neutrinos~\cite{MINOS:2017cae}.
We adopt the analysis procedure followed by the experimental collaboration for the search of active-sterile neutrino oscillations in Ref.~\cite{MINOS:2017cae} by adapting the public MINOS/MINOS+ code to account for LED neutrino oscillations instead of active-sterile oscillations. 

In addition to the data collected by MINOS/MINOS+, we also analyze data from the Daya Bay reactor neutrino experiment~\cite{DayaBay:2018yms}. Daya Bay uses several nuclear reactors summing up a total thermal power of $\sim 17~\text{GW}_{\text{th}}$ and measures the antineutrinos at 8 identical detectors, each with $20~\text{ton}$ fiducial mass, situated at three different sites (experimental halls). For this analysis, we consider the data set corresponding to $1958~\text{days}$~\cite{DayaBay:2018yms}. First, we reproduced the three-neutrino analysis performed by the collaboration using information from Refs.~\cite{DayaBay:2016ssb,DayaBay:2016ggj,DayaBay:2018yms} through an implementation of the experiment in the GLoBES C-library~\cite{Huber:2004ka,Huber:2007ji}. Instead of a far-over-near ratio analysis, the spectral information at the three halls was used. Systematic uncertainties were also included in the analysis, in the same way as in the analysis in Ref.~\cite{deSalas:2020pgw}. After that, we modified GLoBES in order to include the LED oscillation probability in Eq.~(\ref{eq:oscprob}).

\begin{figure}
  \centering
  \includegraphics[width=0.49\textwidth]{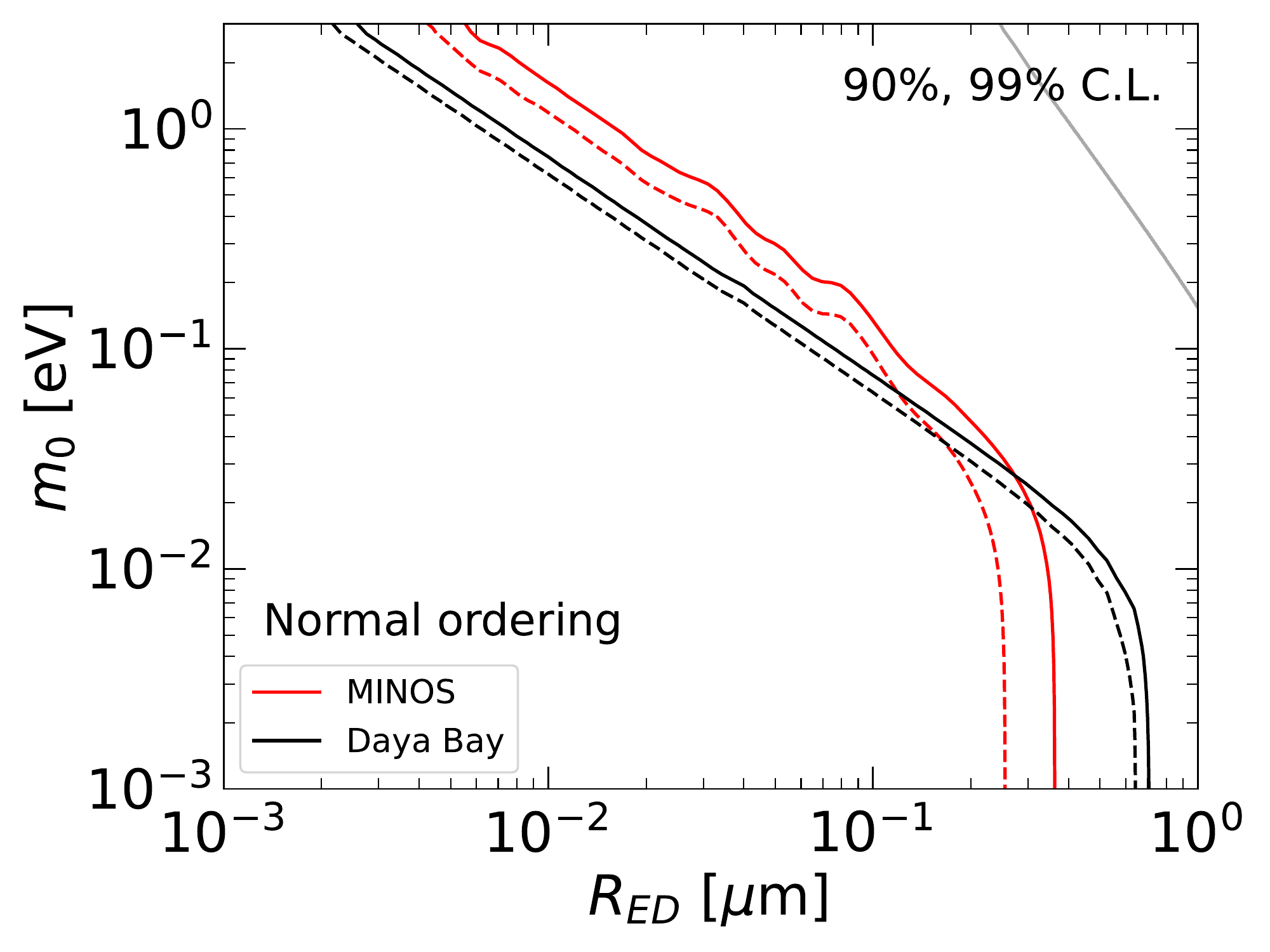}
    \includegraphics[width=0.49\textwidth]{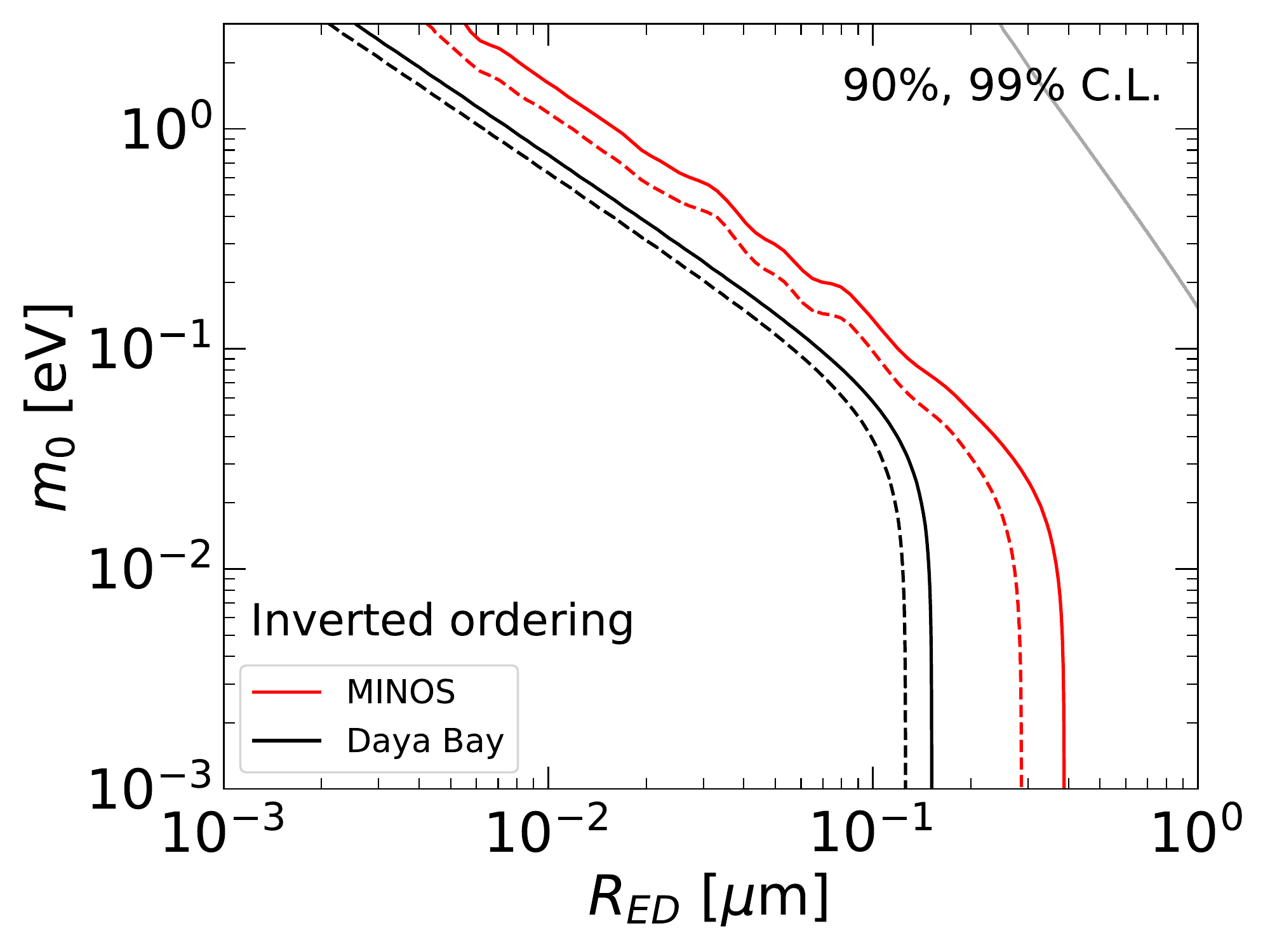}
    \caption{The result from our analysis of data from MINOS/MINOS+ (red) and Daya Bay (black) for normal (left) and inverted (right) neutrino mass ordering. The gray lines in the upper right corner mark the limit of the physically allowed region of parameter space.} 
  \label{fig:lbl}
\end{figure}

The results of our analysis are shown in Fig.~\ref{fig:lbl}. The red and black lines correspond to the bounds at 90\% (dashed) and 99\% (solid) C.L. for two degrees of freedom for MINOS/MINOS+ and Daya Bay, respectively. The results of our analysis of MINOS/MINOS+ data are in reasonable agreement with the preliminary results obtained by the MINOS collaboration in Ref.~\cite{DeRijck:2017wuy}. Note, however, that we perform a simple $\chi^2$ analysis, while the results from Ref.~\cite{DeRijck:2017wuy} are obtained from a Monte Carlo analysis. We find that Daya Bay sets the strongest bounds on the LED parameters for inverted ordering, while for normal ordering the bound by MINOS/MINOS+ is the strongest for small values of $m_0$. It should be noted, that these bounds exclude the LED explanation of the Gallium anomaly and are in even stronger tension with the Gallium region than the regions preferred from the analysis of reactor rate data. It should also be noted that for small values of $R_{\text{ED}}$ the bounds become very weak, since in this case LED effects become very small and then $m_0$ corresponds to the overall neutrino mass scale to which oscillation experiments are not sensitive.

%%%%%%%%%%%%%%%%%%%%%%
\subsection{LED at KATRIN}
\label{sec:kat}
%%%%%%%%%%%%%%%%%%%%%%

In this subsection we present the results of our analysis of the data of the KATRIN experiment. KATRIN is an experiment for direct neutrino mass measurement. It measures the single beta-decay of molecular Tritium, $\text{T}_2\rightarrow {}^3\text{HeT}^+ + e^- + \overline{\nu}_e$, near the endpoint of the spectrum.
The KATRIN collaboration presented their first two mass measurements in Refs.~\cite{KATRIN:2019yun} and~\cite{KATRIN:2021uub}.
From the first (second) campaign,
they obtained an upper bound of 1.1~eV (0.9~eV) at 90\% C.L. on the effective electron neutrino mass $m_{\beta}$
in the standard three-neutrino mixing framework,
which is given by
\begin{equation}
m_{\beta}^2
=
\sum_{i=1}^{3} |U_{ei}|^2 m_i^2
.
\label{m_beta}
\end{equation}
The combined upper bound from both data sets is 0.8~eV at 90\% C.L.~\cite{KATRIN:2021uub}.
A best fit of $m_{\beta}^2=0.26\,\textrm{eV}^2$ was obtained in the second campaign. The KATRIN data has also been used to search for light sterile neutrinos, see Refs.~\cite{Giunti:2019fcj,KATRIN:2020dpx,KATRIN:2022ith}.
Using the current data KATRIN is sensitive to neutrino masses up to 40~eV,
because the data span the last 40 eV of the integral spectrum.

In our analysis we use the KATRIN data given in Ref.~\cite{KATRIN:2022ith} (corresponding to the second campaign) to search for the effects of large extra dimensions. In the presence of LED, the Kurie function is~\cite{Basto-Gonzalez:2012nel}
\begin{equation}\label{eq:Kurie}
    K(E,m_0,R_{ED})=\sum_j p_j \epsilon_j\sum_{i=1}^3|U_{ei}|^2\sum_{n=0}^{\infty}(V_{in})^2\sqrt{\epsilon_j^2-\left(\frac{\lambda_i^{(n)}}{R_{\text{ED}}} \right)^2}\,\Theta\left(\epsilon_j-\frac{\lambda_i^{(n)}}{R_{\text{ED}}}\right)\,
\end{equation}
where $\epsilon_j=E_0-V_j-E$ is the neutrino energy, which depends on the endpoint of the spectrum $E_0$ and the energies of the different final states $V_j$ (which occur with probability $p_j$). In our analysis we include several sources of systematic uncertainties, following the discussion in Ref.~\cite{KATRIN:2022ith}.
In particular, we fit the data with a free total normalization of the spectrum,
a free flat background component,
a hypothetical retarding-potential-dependent
background,
and a small time-dependent background contribution from electrons stored in the Penning trap. Moreover, we constrain the $Q$-value to be
$18575.72 \pm 0.07~\text{eV}$,
as determined from the precise measurement of the atomic mass difference of Tritium and ${}^3$He in Ref.~\cite{Myers:2015lca}.
This $Q$-value implies the endpoint of the spectrum
$E_0 = 18574.21 \pm 0.6~\text{eV}$~\cite{KATRIN:2021uub}.

\begin{figure}
  \centering
  \includegraphics[width=0.49\textwidth]{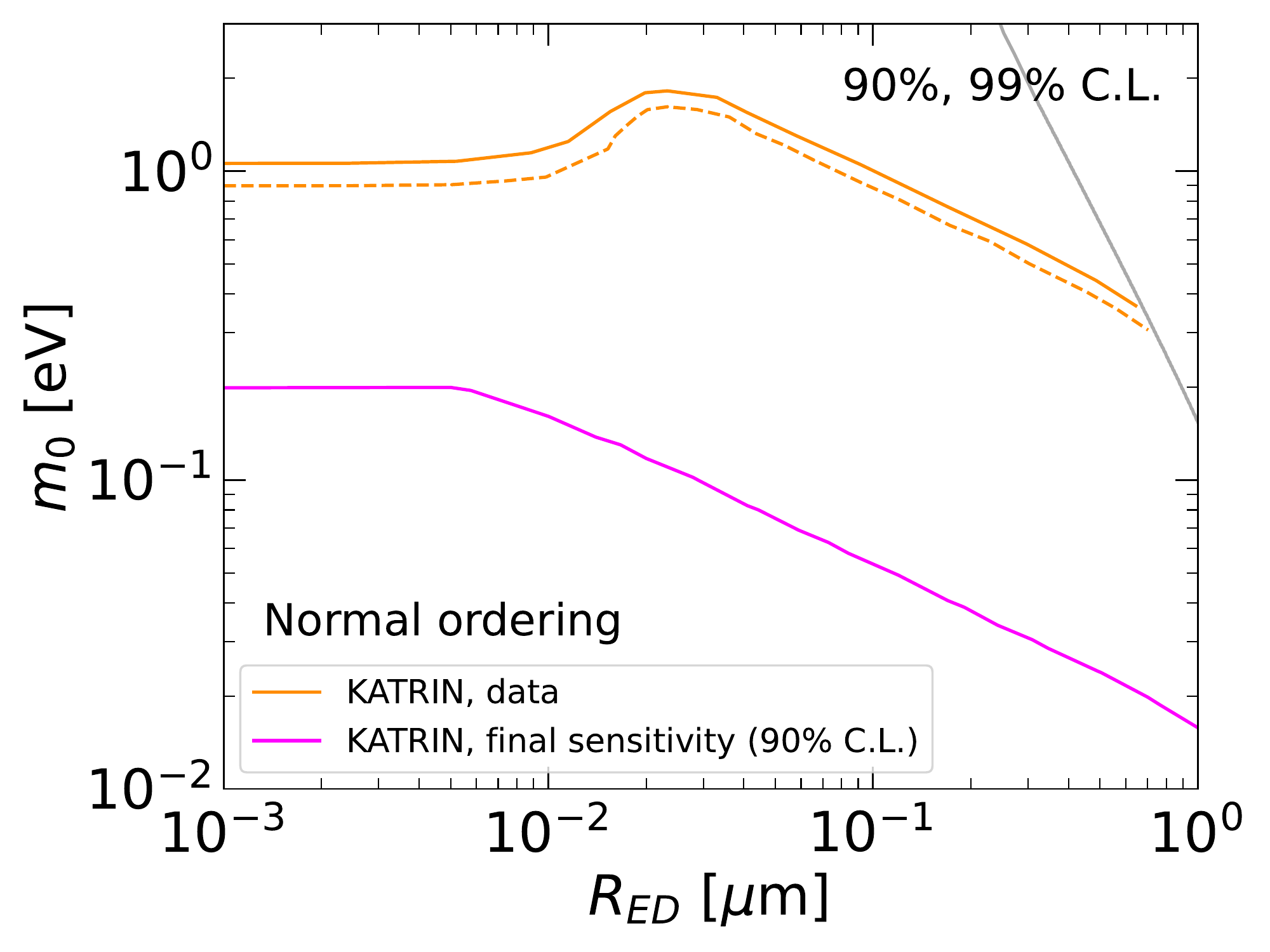}
    \includegraphics[width=0.49\textwidth]{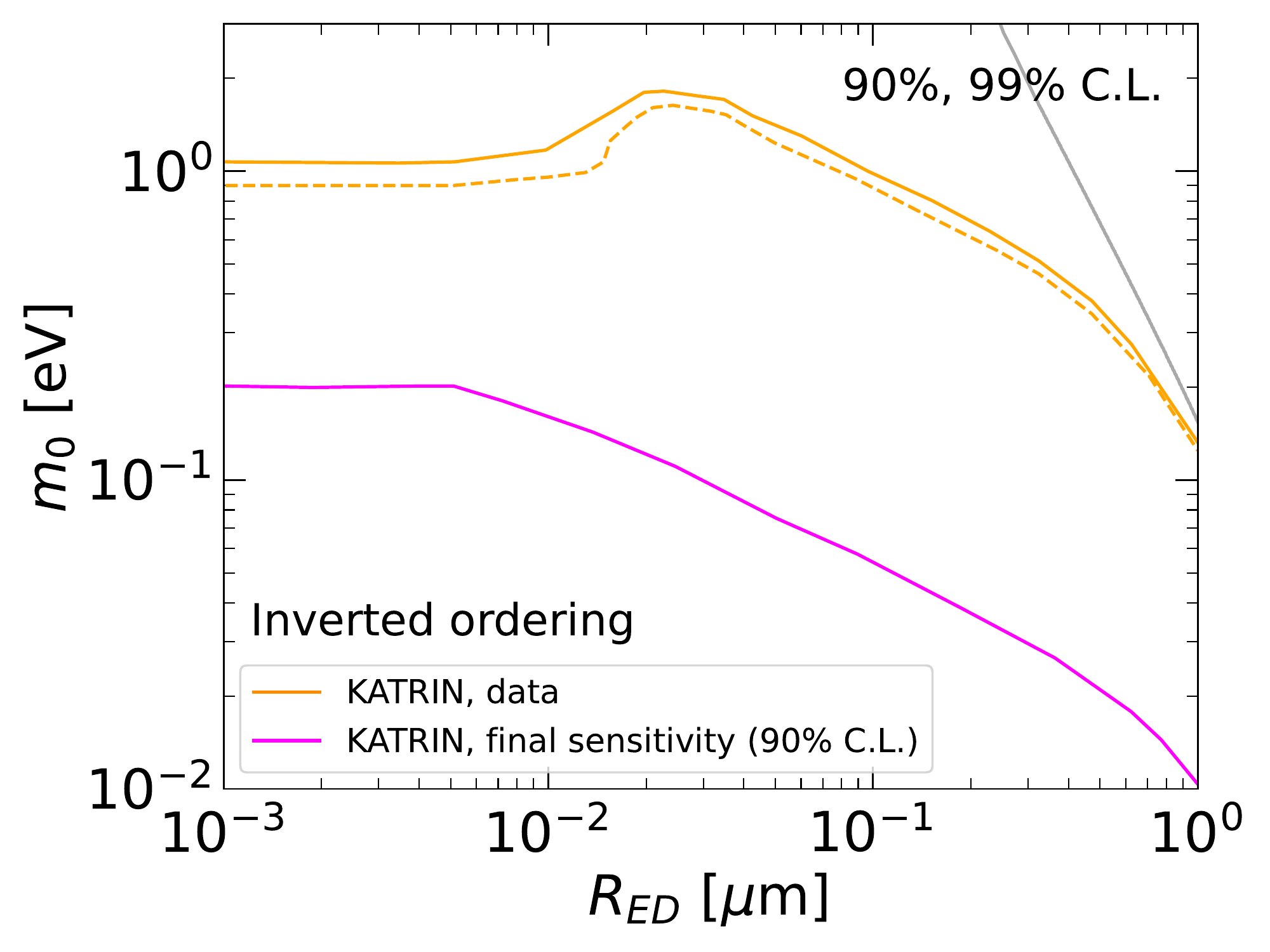}
    \caption{Result from our analysis of data from KATRIN for normal (left) and inverted (right) neutrino mass ordering and also our estimated final sensitivity. Note that in the case of normal ordering the bound stops at the physically allowed limit, indicated by the gray line.} 
  \label{fig:KATRIN}
\end{figure}

We first verified our analysis method by repeating the standard three-neutrino analysis using the data from the second campaign. We obtained an upper bound of $m_{\beta}<0.83$~eV at 90\% C.L. and a best fit $m_{\beta}^2=0.1\,\textrm{eV}^2$. These results are consistent with the results of the KATRIN collaboration
presented in Ref.~\cite{KATRIN:2021uub}.
Next, we exchanged the standard Kurie function with Eq.~\eqref{eq:Kurie} in our analysis code in order to bound the LED parameters. 
In the calculation of the Kurie function all KK modes with masses less than 40~eV have to be included.
The results of our analyses are shown in Fig.~\ref{fig:KATRIN}, where we show the 90\% (dashed orange lines) and 99\% (solid orange lines) C.L. bounds for the LED parameters for normal (left) and inverted (right) neutrino mass ordering.

For
$R_{ED} \lesssim 5\times10^{-3}\,\mu\text{m}$
the upper bound on $m_0$
is the same as the upper bound that we obtained for $m_{\beta}$.
The difference
of the upper bound 0.9 eV at 90\% C.L. in Fig.~\ref{fig:KATRIN},
with the 0.83 eV bound declared above is due to the fact that
the contours in Fig.~\ref{fig:KATRIN} are calculated with two degrees of freedom,
instead of one.
The coincidence of the two bounds is due to the negligible contribution of the KK modes for
$R_{ED} \lesssim 5\times10^{-3}\,\mu\text{m}$.
In this case,
only the zero mode is relevant and $m_0$ is equivalent to $m_{\beta}$.

For
$R_{ED} \gtrsim 5\times10^{-3}\,\mu\text{m}$
the KK modes start to be relevant and
for
$R_{ED}\sim 1\,\mu\text{m}$ more than 200 KK modes must be considered. 

An interesting feature that can be noticed in Fig.~\ref{fig:KATRIN}
is the bump of the bound that occurs around $R_{ED}\sim3\times10^{-2}\,\mu\text{m}$. 
We think that this feature is due to the spacing of the KATRIN data points.
We checked this explanation with a fit of simulated data having a different spacing.
We generated a mock data sample using 100 bins of the retarding potential which are evenly spaced in the 40~eV interval below the endpoint (compared to the 23 slightly unevenly spaced points used by KATRIN), and we found a bound without the bump.
This implies that the bump is caused by the particular retarding potentials at which the KATRIN measurements have been done. A similar bump, in the same mass range, was found in the sterile neutrino analysis of the KATRIN collaboration, see Ref.~\cite{KATRIN:2022ith}. 

Note that the bound on the LED parameters obtained from the analysis of the KATRIN data is not very strong in comparison to the bounds obtained in the previous subsections. It should be noted, however, that the KATRIN bound does not vanish for small values of $R_{\text{ED}}$, where it simply corresponds to the bound on the neutrino mass scale of the standard three-neutrino analysis. Therefore, the analysis of KATRIN data allows us to reduce the volume of the allowed parameter space of the LED parameters, which from the analysis of oscillation data alone is unbound for small values of $R_{\text{ED}}$. In Fig.~\ref{fig:KATRIN} we also show the expected final sensitivity\footnote{Note that our sensitivity estimation is a bit weaker than that of Ref.~\cite{Basto-Gonzalez:2012nel}. The difference is due to the fact that our estimation uses more up-to-date information of the experimental details than the estimation of Ref.~\cite{Basto-Gonzalez:2012nel} done in 2012.} of KATRIN (magenta lines). We see that the KATRIN bound will improve significantly in the near future.

%%%%%%%%%%%%%%%%%%%%%%
%%%%%%%%%%%%%%%%%%%%%%
\section{Combined bounds on LED parameters}
\label{sec:comb}
%%%%%%%%%%%%%%%%%%%%%%
%%%%%%%%%%%%%%%%%%%%%%

In this section we discuss the combined analysis of the experiments considered individually in Section~\ref{sec:data}. We do not consider the Gallium data in this section, since it is in tension with the other data. Regarding the reactor rate data, we consider only the analysis using the KI fluxes. Note that since the combined analysis is dominated by MINOS and Daya Bay, using another model of the reactor antineutrino fluxes would not affect the combined bound on the LED parameters.

\begin{figure}
  \centering
  \includegraphics[width=0.49\textwidth]{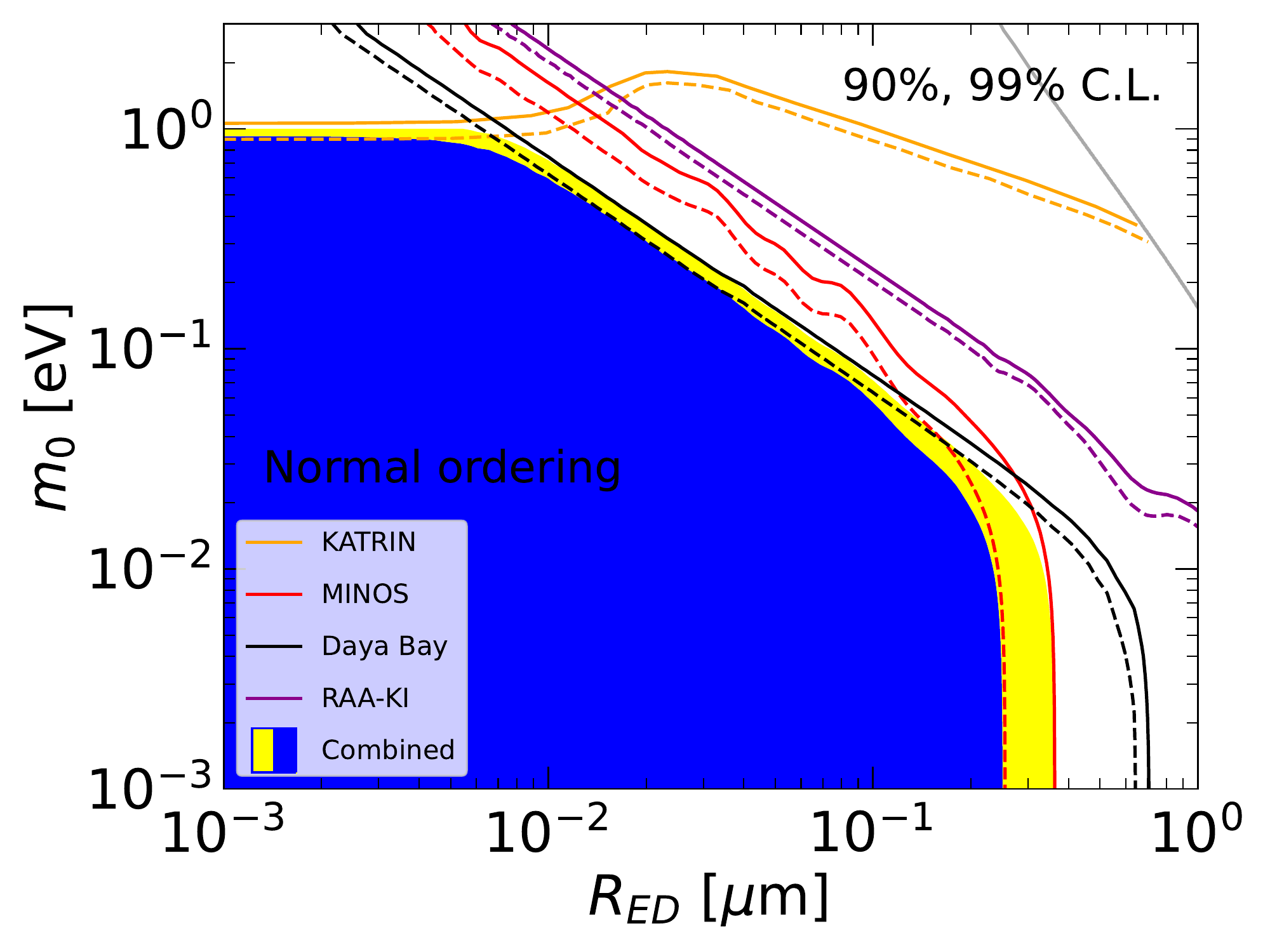}
    \includegraphics[width=0.49\textwidth]{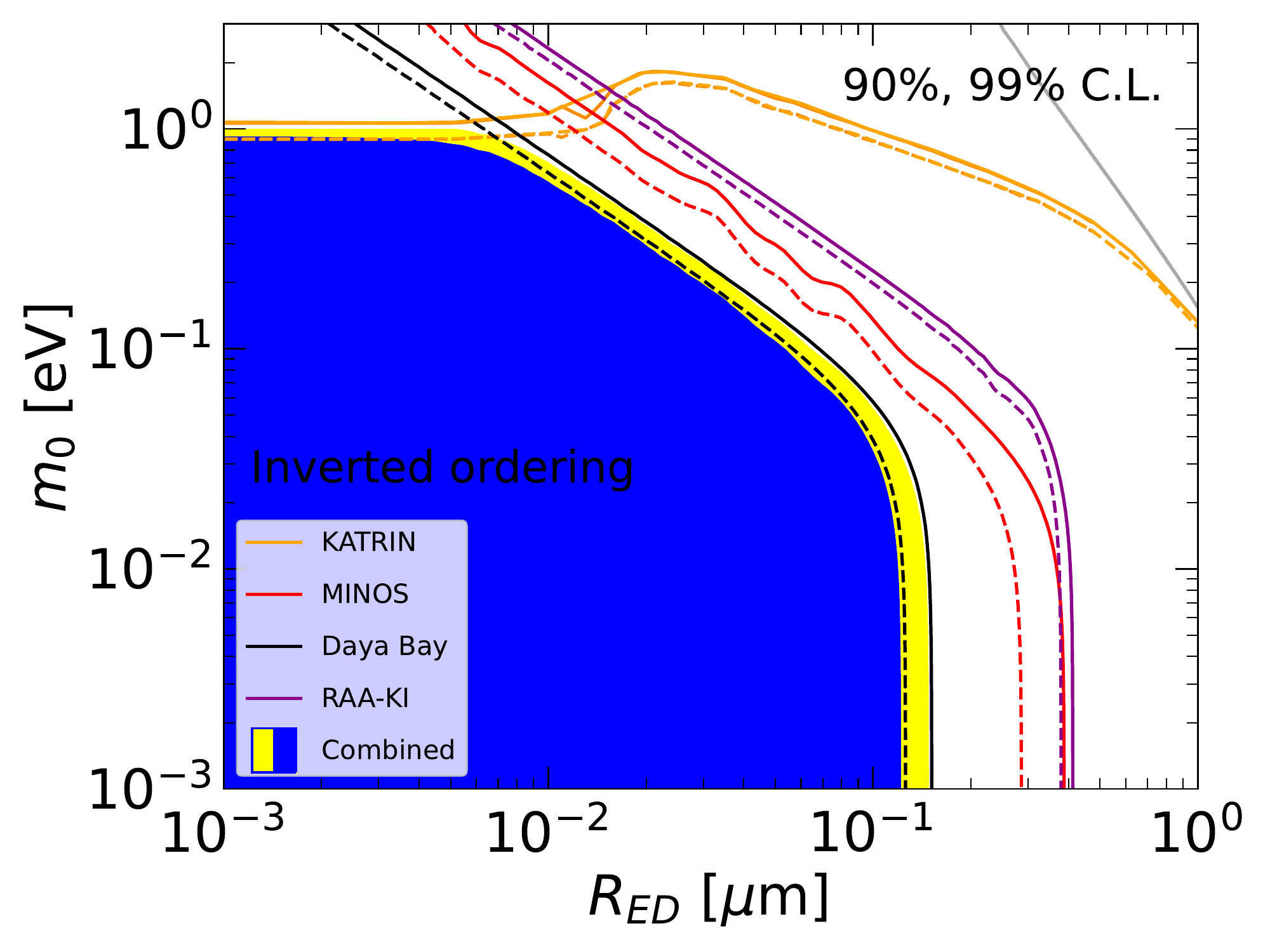}
    \caption{The result from our analysis of data from MINOS/MINOS+ (red) and Daya Bay (black) and the reactor rate data for the KI flux model for normal (left) and inverted (right) neutrino mass ordering. We also show the allowed regions (yellow and blue) from the combined analysis of all data. The gray lines in the upper right corner mark the limit of the physically allowed region of parameter space.} 
  \label{fig:comb}
\end{figure}

\begin{figure}
  \centering
  \includegraphics[width=0.49\textwidth]{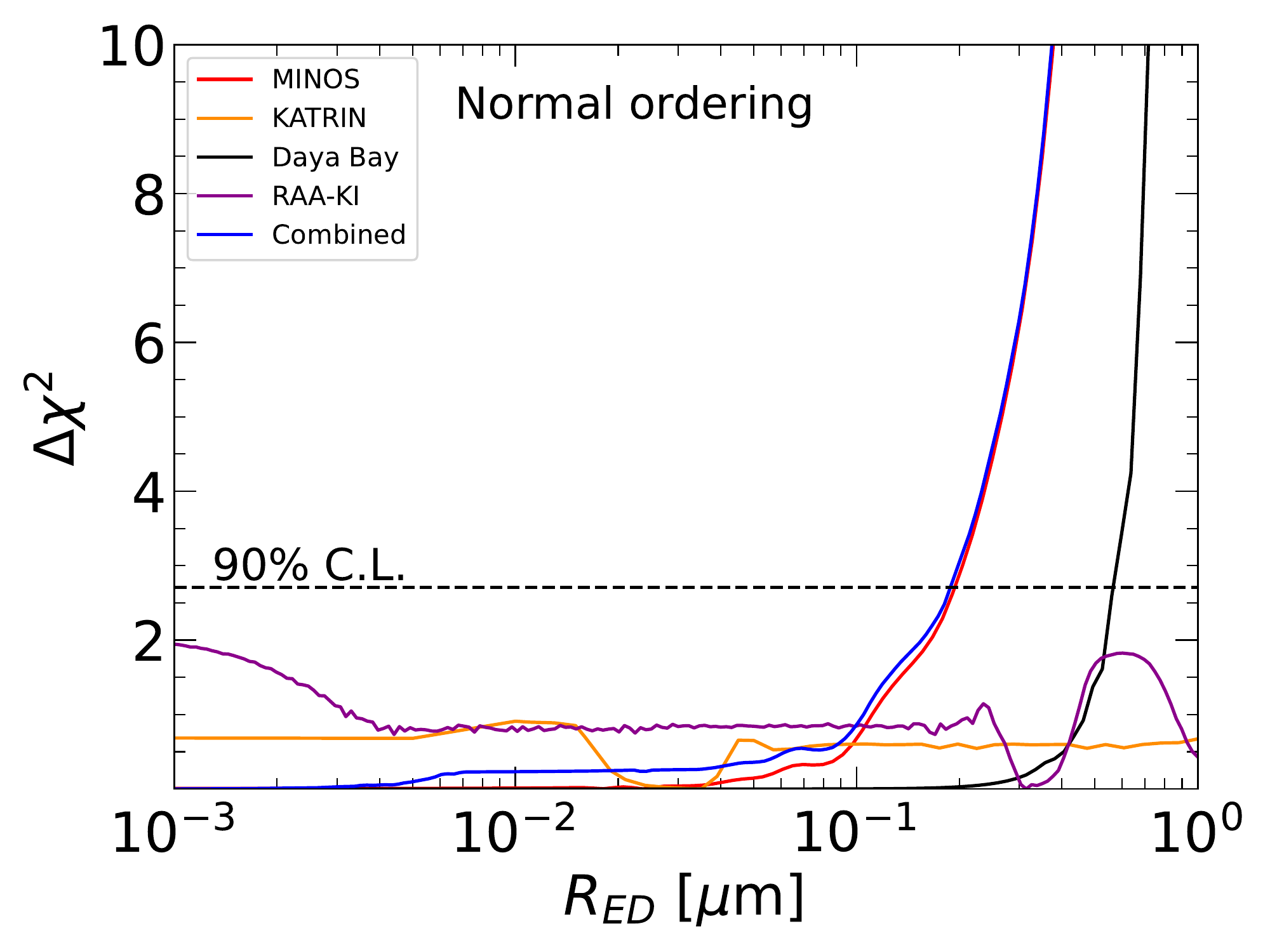}
    \includegraphics[width=0.49\textwidth]{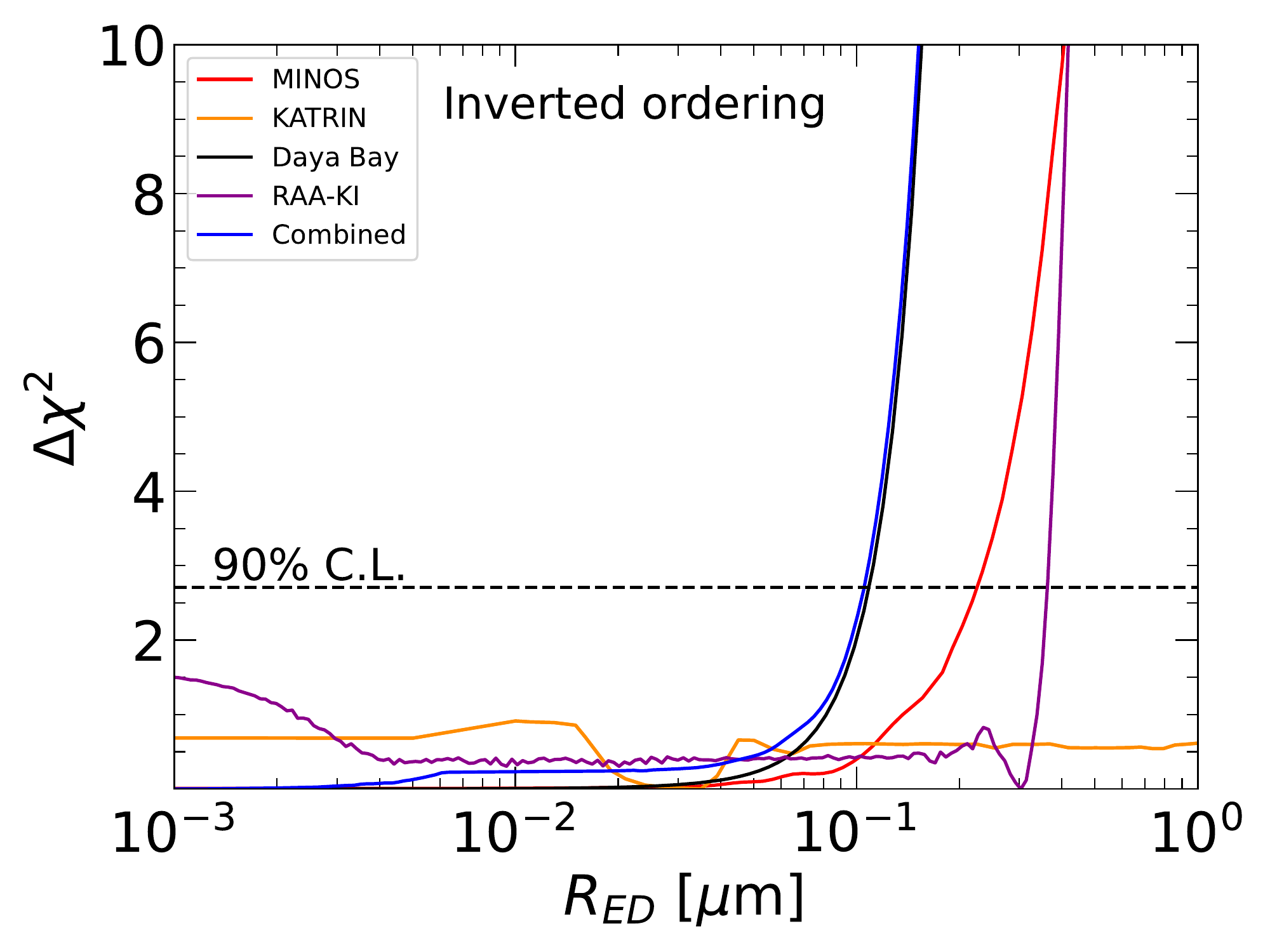}
    \caption{The $\chi^2$ profiles of the radius of the extra dimension $R_{\text{ED}}$ obtained from the analyses discussed in this paper.} 
  \label{fig:comb_R}
\end{figure}

The result of our analysis is shown in Figs.~\ref{fig:comb} and \ref{fig:comb_R}. In Fig.~\ref{fig:comb} we show the 2-dimensional allowed regions for the individual experiments (below the corresponding lines) and the allowed regions obtained from the combined analysis (colored regions). As it can be seen in the right panel of the figure, where the black Daya Bay lines nearly coincide with the boundaries of the blue and yellow regions obtained from the combined analysis, for inverted ordering the combination is dominated by Daya Bay for $R_{\text{ED}} \gtrsim 7 \times 10^{-3}~\mu\text{m}$. Instead, for normal ordering both MINOS and Daya Bay give essential contributions to the definition of the combined allowed region for $R_{\text{ED}} \gtrsim 7 \times 10^{-3}~\mu\text{m}$. As it can be seen, MINOS is dominating for large values of $R_{\text{ED}}$, while Daya Bay is dominating for intermediate values of $R_{\text{ED}}$. For both orderings the contribution of KATRIN is to close the allowed region for small values of $R_{\text{ED}}$. For small values of $R_{\text{ED}}$ the LED oscillation probability becomes basically the standard three-neutrino oscillation probability and then the oscillation experiments can not bound $m_0$, which in this case is simply the neutrino mass scale. In Fig.~\ref{fig:KATRIN} we showed the projected final sensitivity of KATRIN. Comparing it with the bounds in Fig.~\ref{fig:comb}, we see that the contribution of KATRIN to bound the LED parameter space will become much more relevant in the near future.

In Fig.~\ref{fig:comb_R} we show the 1-dimensional $\Delta\chi^2$ profiles for the compactification radius of the large extra dimension. Also from this figure one sees that for normal (inverted) ordering the bound on $R_{\text{ED}}$ is dominated by the analysis of data from MINOS/MINOS+ (Daya Bay). From our combined analyses we find that 
\begin{eqnarray}
    R_{\text{ED}} &<& 0.20~\mu\text{m at 90\% C.L. for NO}\,,\\
    R_{\text{ED}} &<& 0.10~\mu\text{m at 90\% C.L. for IO}\,.
\end{eqnarray}
These bounds are quite strong for both orderings.
In spite of the potentialities of the next generation of neutrino oscillation experiments,
unfortunately it is unlikely that they will be able to improve these bounds
according to Ref.~\cite{Basto-Gonzalez:2021aus} for JUNO and TAO,
Refs.~\cite{Berryman:2016szd,Arguelles:2019xgp,DUNE:2020fgq} for DUNE,
and
Ref.~\cite{Stenico:2018jpl} for SBN.

Let us finally emphasize that the results of our analysis improve all the bounds on the LED parameters obtained previously from neutrino oscillation data~\cite{Davoudiasl:2002fq,Machado:2011jt,MINOS:2016vvv,Esmaili:2014esa,Girardi:2014gna,DiIura:2014csa}.

%%%%%%%%%%%%%%%%%%%%%%
%%%%%%%%%%%%%%%%%%%%%%
\section{Conclusions}
\label{sec:conc}
%%%%%%%%%%%%%%%%%%%%%%
%%%%%%%%%%%%%%%%%%%%%%
We have performed an analysis of the data of several neutrino oscillation experiments and the KATRIN experiment in the framework of a Large Extra Dimensions model with a dominant large extra dimension.
We obtained the strongest bounds on the LED parameter space that have been attained so far from the data of neutrino experiments. We have shown that the LED explanation of the Gallium anomaly is excluded by the analyses of the data of the other experiments. This behavior has also been observed in the context of neutrino mixing with a light sterile neutrino~\cite{Giunti:2021kab}. Combining all data we obtained a bound of $R_{\text{ED}} < 0.20 \ (0.10)~\mu\text{m}$ at 90\% C.L. for NO (IO). Using KATRIN data we are able to close the region at small values of $R_{\text{ED}}$, as it can be seen in Fig.~\ref{fig:comb}. The KATRIN bound on $m_0$ is expected to improve by approximately one order of magnitude, according to the final KATRIN sensitivity. Hence, the role of KATRIN will become more crucial in the future.

Apart from neutrino experiments, searches for Large Extra Dimensions have been performed using tabletop gravitational experiments~\cite{Long:1998dk,Krause:1999ry,Fischbach:2001ry,Adelberger:2002ic},
collider experiments~\cite{Rizzo:1998fm,Hewett:1998sn,D0:2000cve,DELPHI:2000ztm,DELPHI:2008uka},
and also performing analyses of astrophysical~\cite{Cullen:1999hc,Barger:1999jf,Hanhart:2000er,Hannestad:2001jv,Hannestad:2003yd} and cosmological data~\cite{Hall:1999mk,Hannestad:2001nq,Fairbairn:2001ct}. Using astrophysical data, very strong constraints ranging in $R_{\text{ED}}<0.16-916\,\text{nm}$ have been obtained. However, these limits depend on the technique and some assumptions~\cite{ParticleDataGroup:2020ssz} in the analyses.
It should be noted that the bound that we obtained from the data of neutrino experiments is two orders of magnitude stronger than the constraints obtained using tabletop experiments, which achieved a limit of $R_{\text{ED}}<37~\mu \text{m}$ at 95\% C.L.~\cite{ParticleDataGroup:2020ssz}.

\section*{Acknowledgments}
We would like to thank
Thierry Lasserre and Lisa Schl\"uter
for providing useful information on the KATRIN data,
and
Sanshiro Enomoto,
Leonard K\"ollenberger, and
Alexey Lokhov
for useful discussions on the KATRIN data analysis at the NuMass 2022 workshop.
C.G. and C.A.T. are supported by the research grant ``The Dark Universe: A Synergic Multimessenger Approach'' number 2017X7X85K under the program ``PRIN 2017'' funded by the Italian Ministero dell'Istruzione, Universit\`a e della Ricerca (MIUR).
O.T. is supported by a grant funded by Italian Ministero degli Affari Esteri e della Cooperazione Internazionale (MAECI) and also by the Indian Prime Minister's Research Fellow (PMRF) program.

%\bibliographystyle{apsrev4-1}
%\bibliography{bibliography}  

\begin{thebibliography}{89}%
\makeatletter
\providecommand \@ifxundefined [1]{%
 \@ifx{#1\undefined}
}%
\providecommand \@ifnum [1]{%
 \ifnum #1\expandafter \@firstoftwo
 \else \expandafter \@secondoftwo
 \fi
}%
\providecommand \@ifx [1]{%
 \ifx #1\expandafter \@firstoftwo
 \else \expandafter \@secondoftwo
 \fi
}%
\providecommand \natexlab [1]{#1}%
\providecommand \enquote  [1]{``#1''}%
\providecommand \bibnamefont  [1]{#1}%
\providecommand \bibfnamefont [1]{#1}%
\providecommand \citenamefont [1]{#1}%
\providecommand \href@noop [0]{\@secondoftwo}%
\providecommand \href [0]{\begingroup \@sanitize@url \@href}%
\providecommand \@href[1]{\@@startlink{#1}\@@href}%
\providecommand \@@href[1]{\endgroup#1\@@endlink}%
\providecommand \@sanitize@url [0]{\catcode `\\12\catcode `\$12\catcode
  `\&12\catcode `\#12\catcode `\^12\catcode `\_12\catcode `\%12\relax}%
\providecommand \@@startlink[1]{}%
\providecommand \@@endlink[0]{}%
\providecommand \url  [0]{\begingroup\@sanitize@url \@url }%
\providecommand \@url [1]{\endgroup\@href {#1}{\urlprefix }}%
\providecommand \urlprefix  [0]{URL }%
\providecommand \Eprint [0]{\href }%
\providecommand \doibase [0]{http://dx.doi.org/}%
\providecommand \selectlanguage [0]{\@gobble}%
\providecommand \bibinfo  [0]{\@secondoftwo}%
\providecommand \bibfield  [0]{\@secondoftwo}%
\providecommand \translation [1]{[#1]}%
\providecommand \BibitemOpen [0]{}%
\providecommand \bibitemStop [0]{}%
\providecommand \bibitemNoStop [0]{.\EOS\space}%
\providecommand \EOS [0]{\spacefactor3000\relax}%
\providecommand \BibitemShut  [1]{\csname bibitem#1\endcsname}%
\let\auto@bib@innerbib\@empty
%</preamble>
\bibitem [{\citenamefont {Mohapatra}\ and\ \citenamefont
  {Smirnov}(2006)}]{Mohapatra:2006gs}%
  \BibitemOpen
  \bibfield  {author} {\bibinfo {author} {\bibfnamefont {R.~N.}\ \bibnamefont
  {Mohapatra}}\ and\ \bibinfo {author} {\bibfnamefont {A.~Y.}\ \bibnamefont
  {Smirnov}},\ }\href {\doibase 10.1146/annurev.nucl.56.080805.140534}
  {\bibfield  {journal} {\bibinfo  {journal} {Ann. Rev. Nucl. Part. Sci.}\
  }\textbf {\bibinfo {volume} {56}},\ \bibinfo {pages} {569} (\bibinfo {year}
  {2006})},\ \Eprint {http://arxiv.org/abs/hep-ph/0603118}
  {arXiv:hep-ph/0603118} \BibitemShut {NoStop}%
\bibitem [{\citenamefont {Gonzalez-Garcia}\ and\ \citenamefont
  {Maltoni}(2008)}]{Gonzalez-Garcia:2007dlo}%
  \BibitemOpen
  \bibfield  {author} {\bibinfo {author} {\bibfnamefont {M.~C.}\ \bibnamefont
  {Gonzalez-Garcia}}\ and\ \bibinfo {author} {\bibfnamefont {M.}~\bibnamefont
  {Maltoni}},\ }\href {\doibase 10.1016/j.physrep.2007.12.004} {\bibfield
  {journal} {\bibinfo  {journal} {Phys. Rept.}\ }\textbf {\bibinfo {volume}
  {460}},\ \bibinfo {pages} {1} (\bibinfo {year} {2008})},\ \Eprint
  {http://arxiv.org/abs/0704.1800} {arXiv:0704.1800 [hep-ph]} \BibitemShut
  {NoStop}%
\bibitem [{\citenamefont {Petcov}(2013)}]{Petcov:2013poa}%
  \BibitemOpen
  \bibfield  {author} {\bibinfo {author} {\bibfnamefont {S.~T.}\ \bibnamefont
  {Petcov}},\ }\href {\doibase 10.1155/2013/852987} {\bibfield  {journal}
  {\bibinfo  {journal} {Adv. High Energy Phys.}\ }\textbf {\bibinfo {volume}
  {2013}},\ \bibinfo {pages} {852987} (\bibinfo {year} {2013})},\ \Eprint
  {http://arxiv.org/abs/1303.5819} {arXiv:1303.5819 [hep-ph]} \BibitemShut
  {NoStop}%
\bibitem [{\citenamefont {King}\ \emph {et~al.}(2014)\citenamefont {King},
  \citenamefont {Merle}, \citenamefont {Morisi}, \citenamefont {Shimizu},\ and\
  \citenamefont {Tanimoto}}]{King:2014nza}%
  \BibitemOpen
  \bibfield  {author} {\bibinfo {author} {\bibfnamefont {S.~F.}\ \bibnamefont
  {King}}, \bibinfo {author} {\bibfnamefont {A.}~\bibnamefont {Merle}},
  \bibinfo {author} {\bibfnamefont {S.}~\bibnamefont {Morisi}}, \bibinfo
  {author} {\bibfnamefont {Y.}~\bibnamefont {Shimizu}}, \ and\ \bibinfo
  {author} {\bibfnamefont {M.}~\bibnamefont {Tanimoto}},\ }\href {\doibase
  10.1088/1367-2630/16/4/045018} {\bibfield  {journal} {\bibinfo  {journal}
  {New J. Phys.}\ }\textbf {\bibinfo {volume} {16}},\ \bibinfo {pages} {045018}
  (\bibinfo {year} {2014})},\ \Eprint {http://arxiv.org/abs/1402.4271}
  {arXiv:1402.4271 [hep-ph]} \BibitemShut {NoStop}%
\bibitem [{\citenamefont {Arkani-Hamed}\ \emph {et~al.}(1998)\citenamefont
  {Arkani-Hamed}, \citenamefont {Dimopoulos},\ and\ \citenamefont
  {Dvali}}]{Arkani-Hamed:1998jmv}%
  \BibitemOpen
  \bibfield  {author} {\bibinfo {author} {\bibfnamefont {N.}~\bibnamefont
  {Arkani-Hamed}}, \bibinfo {author} {\bibfnamefont {S.}~\bibnamefont
  {Dimopoulos}}, \ and\ \bibinfo {author} {\bibfnamefont {G.~R.}\ \bibnamefont
  {Dvali}},\ }\href {\doibase 10.1016/S0370-2693(98)00466-3} {\bibfield
  {journal} {\bibinfo  {journal} {Phys. Lett. B}\ }\textbf {\bibinfo {volume}
  {429}},\ \bibinfo {pages} {263} (\bibinfo {year} {1998})},\ \Eprint
  {http://arxiv.org/abs/hep-ph/9803315} {arXiv:hep-ph/9803315} \BibitemShut
  {NoStop}%
\bibitem [{\citenamefont {Arkani-Hamed}\ \emph {et~al.}(1999)\citenamefont
  {Arkani-Hamed}, \citenamefont {Dimopoulos},\ and\ \citenamefont
  {Dvali}}]{Arkani-Hamed:1998sfv}%
  \BibitemOpen
  \bibfield  {author} {\bibinfo {author} {\bibfnamefont {N.}~\bibnamefont
  {Arkani-Hamed}}, \bibinfo {author} {\bibfnamefont {S.}~\bibnamefont
  {Dimopoulos}}, \ and\ \bibinfo {author} {\bibfnamefont {G.~R.}\ \bibnamefont
  {Dvali}},\ }\href {\doibase 10.1103/PhysRevD.59.086004} {\bibfield  {journal}
  {\bibinfo  {journal} {Phys. Rev. D}\ }\textbf {\bibinfo {volume} {59}},\
  \bibinfo {pages} {086004} (\bibinfo {year} {1999})},\ \Eprint
  {http://arxiv.org/abs/hep-ph/9807344} {arXiv:hep-ph/9807344} \BibitemShut
  {NoStop}%
\bibitem [{\citenamefont {Arkani-Hamed}\ \emph {et~al.}(2001)\citenamefont
  {Arkani-Hamed}, \citenamefont {Dimopoulos}, \citenamefont {Dvali},\ and\
  \citenamefont {March-Russell}}]{Arkani-Hamed:1998wuz}%
  \BibitemOpen
  \bibfield  {author} {\bibinfo {author} {\bibfnamefont {N.}~\bibnamefont
  {Arkani-Hamed}}, \bibinfo {author} {\bibfnamefont {S.}~\bibnamefont
  {Dimopoulos}}, \bibinfo {author} {\bibfnamefont {G.~R.}\ \bibnamefont
  {Dvali}}, \ and\ \bibinfo {author} {\bibfnamefont {J.}~\bibnamefont
  {March-Russell}},\ }\href {\doibase 10.1103/PhysRevD.65.024032} {\bibfield
  {journal} {\bibinfo  {journal} {Phys. Rev. D}\ }\textbf {\bibinfo {volume}
  {65}},\ \bibinfo {pages} {024032} (\bibinfo {year} {2001})},\ \Eprint
  {http://arxiv.org/abs/hep-ph/9811448} {arXiv:hep-ph/9811448} \BibitemShut
  {NoStop}%
\bibitem [{\citenamefont {Dienes}\ \emph {et~al.}(1999)\citenamefont {Dienes},
  \citenamefont {Dudas},\ and\ \citenamefont {Gherghetta}}]{Dienes:1998sb}%
  \BibitemOpen
  \bibfield  {author} {\bibinfo {author} {\bibfnamefont {K.~R.}\ \bibnamefont
  {Dienes}}, \bibinfo {author} {\bibfnamefont {E.}~\bibnamefont {Dudas}}, \
  and\ \bibinfo {author} {\bibfnamefont {T.}~\bibnamefont {Gherghetta}},\
  }\href {\doibase 10.1016/S0550-3213(99)00377-6} {\bibfield  {journal}
  {\bibinfo  {journal} {Nucl. Phys. B}\ }\textbf {\bibinfo {volume} {557}},\
  \bibinfo {pages} {25} (\bibinfo {year} {1999})},\ \Eprint
  {http://arxiv.org/abs/hep-ph/9811428} {arXiv:hep-ph/9811428} \BibitemShut
  {NoStop}%
\bibitem [{\citenamefont {Dvali}\ and\ \citenamefont
  {Smirnov}(1999)}]{Dvali:1999cn}%
  \BibitemOpen
  \bibfield  {author} {\bibinfo {author} {\bibfnamefont {G.~R.}\ \bibnamefont
  {Dvali}}\ and\ \bibinfo {author} {\bibfnamefont {A.~Y.}\ \bibnamefont
  {Smirnov}},\ }\href {\doibase 10.1016/S0550-3213(99)00574-X} {\bibfield
  {journal} {\bibinfo  {journal} {Nucl. Phys. B}\ }\textbf {\bibinfo {volume}
  {563}},\ \bibinfo {pages} {63} (\bibinfo {year} {1999})},\ \Eprint
  {http://arxiv.org/abs/hep-ph/9904211} {arXiv:hep-ph/9904211} \BibitemShut
  {NoStop}%
\bibitem [{\citenamefont {Mohapatra}\ and\ \citenamefont
  {Perez-Lorenzana}(2001)}]{Mohapatra:2000wn}%
  \BibitemOpen
  \bibfield  {author} {\bibinfo {author} {\bibfnamefont {R.~N.}\ \bibnamefont
  {Mohapatra}}\ and\ \bibinfo {author} {\bibfnamefont {A.}~\bibnamefont
  {Perez-Lorenzana}},\ }\href {\doibase 10.1016/S0550-3213(00)00634-9}
  {\bibfield  {journal} {\bibinfo  {journal} {Nucl. Phys. B}\ }\textbf
  {\bibinfo {volume} {593}},\ \bibinfo {pages} {451} (\bibinfo {year}
  {2001})},\ \Eprint {http://arxiv.org/abs/hep-ph/0006278}
  {arXiv:hep-ph/0006278} \BibitemShut {NoStop}%
\bibitem [{\citenamefont {Barbieri}\ \emph {et~al.}(2000)\citenamefont
  {Barbieri}, \citenamefont {Creminelli},\ and\ \citenamefont
  {Strumia}}]{Barbieri:2000mg}%
  \BibitemOpen
  \bibfield  {author} {\bibinfo {author} {\bibfnamefont {R.}~\bibnamefont
  {Barbieri}}, \bibinfo {author} {\bibfnamefont {P.}~\bibnamefont
  {Creminelli}}, \ and\ \bibinfo {author} {\bibfnamefont {A.}~\bibnamefont
  {Strumia}},\ }\href {\doibase 10.1016/S0550-3213(00)00348-5} {\bibfield
  {journal} {\bibinfo  {journal} {Nucl. Phys. B}\ }\textbf {\bibinfo {volume}
  {585}},\ \bibinfo {pages} {28} (\bibinfo {year} {2000})},\ \Eprint
  {http://arxiv.org/abs/hep-ph/0002199} {arXiv:hep-ph/0002199} \BibitemShut
  {NoStop}%
\bibitem [{\citenamefont {Davoudiasl}\ \emph {et~al.}(2002)\citenamefont
  {Davoudiasl}, \citenamefont {Langacker},\ and\ \citenamefont
  {Perelstein}}]{Davoudiasl:2002fq}%
  \BibitemOpen
  \bibfield  {author} {\bibinfo {author} {\bibfnamefont {H.}~\bibnamefont
  {Davoudiasl}}, \bibinfo {author} {\bibfnamefont {P.}~\bibnamefont
  {Langacker}}, \ and\ \bibinfo {author} {\bibfnamefont {M.}~\bibnamefont
  {Perelstein}},\ }\href {\doibase 10.1103/PhysRevD.65.105015} {\bibfield
  {journal} {\bibinfo  {journal} {Phys. Rev. D}\ }\textbf {\bibinfo {volume}
  {65}},\ \bibinfo {pages} {105015} (\bibinfo {year} {2002})},\ \Eprint
  {http://arxiv.org/abs/hep-ph/0201128} {arXiv:hep-ph/0201128} \BibitemShut
  {NoStop}%
\bibitem [{\citenamefont {Esmaili}\ \emph {et~al.}(2014)\citenamefont
  {Esmaili}, \citenamefont {Peres},\ and\ \citenamefont
  {Tabrizi}}]{Esmaili:2014esa}%
  \BibitemOpen
  \bibfield  {author} {\bibinfo {author} {\bibfnamefont {A.}~\bibnamefont
  {Esmaili}}, \bibinfo {author} {\bibfnamefont {O.~L.~G.}\ \bibnamefont
  {Peres}}, \ and\ \bibinfo {author} {\bibfnamefont {Z.}~\bibnamefont
  {Tabrizi}},\ }\href {\doibase 10.1088/1475-7516/2014/12/002} {\bibfield
  {journal} {\bibinfo  {journal} {JCAP}\ }\textbf {\bibinfo {volume} {12}},\
  \bibinfo {pages} {002} (\bibinfo {year} {2014})},\ \Eprint
  {http://arxiv.org/abs/1409.3502} {arXiv:1409.3502 [hep-ph]} \BibitemShut
  {NoStop}%
\bibitem [{\citenamefont {de~Salas}\ \emph {et~al.}(2021)\citenamefont
  {de~Salas}, \citenamefont {Forero}, \citenamefont {Gariazzo}, \citenamefont
  {Mart\'\i{}nez-Mirav\'e}, \citenamefont {Mena}, \citenamefont {Ternes},
  \citenamefont {T\'ortola},\ and\ \citenamefont {Valle}}]{deSalas:2020pgw}%
  \BibitemOpen
  \bibfield  {author} {\bibinfo {author} {\bibfnamefont {P.~F.}\ \bibnamefont
  {de~Salas}}, \bibinfo {author} {\bibfnamefont {D.~V.}\ \bibnamefont
  {Forero}}, \bibinfo {author} {\bibfnamefont {S.}~\bibnamefont {Gariazzo}},
  \bibinfo {author} {\bibfnamefont {P.}~\bibnamefont {Mart\'\i{}nez-Mirav\'e}},
  \bibinfo {author} {\bibfnamefont {O.}~\bibnamefont {Mena}}, \bibinfo {author}
  {\bibfnamefont {C.~A.}\ \bibnamefont {Ternes}}, \bibinfo {author}
  {\bibfnamefont {M.}~\bibnamefont {T\'ortola}}, \ and\ \bibinfo {author}
  {\bibfnamefont {J.~W.~F.}\ \bibnamefont {Valle}},\ }\href {\doibase
  10.1007/JHEP02(2021)071} {\bibfield  {journal} {\bibinfo  {journal} {JHEP}\
  }\textbf {\bibinfo {volume} {02}},\ \bibinfo {pages} {071} (\bibinfo {year}
  {2021})},\ \Eprint {http://arxiv.org/abs/2006.11237} {arXiv:2006.11237
  [hep-ph]} \BibitemShut {NoStop}%
\bibitem [{\citenamefont {Esteban}\ \emph {et~al.}(2020)\citenamefont
  {Esteban}, \citenamefont {Gonzalez-Garcia}, \citenamefont {Maltoni},
  \citenamefont {Schwetz},\ and\ \citenamefont {Zhou}}]{Esteban:2020cvm}%
  \BibitemOpen
  \bibfield  {author} {\bibinfo {author} {\bibfnamefont {I.}~\bibnamefont
  {Esteban}}, \bibinfo {author} {\bibfnamefont {M.~C.}\ \bibnamefont
  {Gonzalez-Garcia}}, \bibinfo {author} {\bibfnamefont {M.}~\bibnamefont
  {Maltoni}}, \bibinfo {author} {\bibfnamefont {T.}~\bibnamefont {Schwetz}}, \
  and\ \bibinfo {author} {\bibfnamefont {A.}~\bibnamefont {Zhou}},\ }\href
  {\doibase 10.1007/JHEP09(2020)178} {\bibfield  {journal} {\bibinfo  {journal}
  {JHEP}\ }\textbf {\bibinfo {volume} {09}},\ \bibinfo {pages} {178} (\bibinfo
  {year} {2020})},\ \Eprint {http://arxiv.org/abs/2007.14792} {arXiv:2007.14792
  [hep-ph]} \BibitemShut {NoStop}%
\bibitem [{\citenamefont {Capozzi}\ \emph {et~al.}(2021)\citenamefont
  {Capozzi}, \citenamefont {Di~Valentino}, \citenamefont {Lisi}, \citenamefont
  {Marrone}, \citenamefont {Melchiorri},\ and\ \citenamefont
  {Palazzo}}]{Capozzi:2021fjo}%
  \BibitemOpen
  \bibfield  {author} {\bibinfo {author} {\bibfnamefont {F.}~\bibnamefont
  {Capozzi}}, \bibinfo {author} {\bibfnamefont {E.}~\bibnamefont
  {Di~Valentino}}, \bibinfo {author} {\bibfnamefont {E.}~\bibnamefont {Lisi}},
  \bibinfo {author} {\bibfnamefont {A.}~\bibnamefont {Marrone}}, \bibinfo
  {author} {\bibfnamefont {A.}~\bibnamefont {Melchiorri}}, \ and\ \bibinfo
  {author} {\bibfnamefont {A.}~\bibnamefont {Palazzo}},\ }\href {\doibase
  10.1103/PhysRevD.104.083031} {\bibfield  {journal} {\bibinfo  {journal}
  {Phys. Rev. D}\ }\textbf {\bibinfo {volume} {104}},\ \bibinfo {pages}
  {083031} (\bibinfo {year} {2021})},\ \Eprint
  {http://arxiv.org/abs/2107.00532} {arXiv:2107.00532 [hep-ph]} \BibitemShut
  {NoStop}%
\bibitem [{\citenamefont {Machado}\ \emph {et~al.}(2011)\citenamefont
  {Machado}, \citenamefont {Nunokawa},\ and\ \citenamefont
  {Zukanovich~Funchal}}]{Machado:2011jt}%
  \BibitemOpen
  \bibfield  {author} {\bibinfo {author} {\bibfnamefont {P.~A.~N.}\
  \bibnamefont {Machado}}, \bibinfo {author} {\bibfnamefont {H.}~\bibnamefont
  {Nunokawa}}, \ and\ \bibinfo {author} {\bibfnamefont {R.}~\bibnamefont
  {Zukanovich~Funchal}},\ }\href {\doibase 10.1103/PhysRevD.84.013003}
  {\bibfield  {journal} {\bibinfo  {journal} {Phys. Rev. D}\ }\textbf {\bibinfo
  {volume} {84}},\ \bibinfo {pages} {013003} (\bibinfo {year} {2011})},\
  \Eprint {http://arxiv.org/abs/1101.0003} {arXiv:1101.0003 [hep-ph]}
  \BibitemShut {NoStop}%
\bibitem [{\citenamefont {Machado}\ \emph {et~al.}(2012)\citenamefont
  {Machado}, \citenamefont {Nunokawa}, \citenamefont {dos Santos},\ and\
  \citenamefont {Funchal}}]{Machado:2011kt}%
  \BibitemOpen
  \bibfield  {author} {\bibinfo {author} {\bibfnamefont {P.~A.~N.}\
  \bibnamefont {Machado}}, \bibinfo {author} {\bibfnamefont {H.}~\bibnamefont
  {Nunokawa}}, \bibinfo {author} {\bibfnamefont {F.~A.~P.}\ \bibnamefont {dos
  Santos}}, \ and\ \bibinfo {author} {\bibfnamefont {R.~Z.}\ \bibnamefont
  {Funchal}},\ }\href {\doibase 10.1103/PhysRevD.85.073012} {\bibfield
  {journal} {\bibinfo  {journal} {Phys. Rev. D}\ }\textbf {\bibinfo {volume}
  {85}},\ \bibinfo {pages} {073012} (\bibinfo {year} {2012})},\ \Eprint
  {http://arxiv.org/abs/1107.2400} {arXiv:1107.2400 [hep-ph]} \BibitemShut
  {NoStop}%
\bibitem [{\citenamefont {Basto-Gonzalez}\ \emph {et~al.}(2013)\citenamefont
  {Basto-Gonzalez}, \citenamefont {Esmaili},\ and\ \citenamefont
  {Peres}}]{Basto-Gonzalez:2012nel}%
  \BibitemOpen
  \bibfield  {author} {\bibinfo {author} {\bibfnamefont {V.~S.}\ \bibnamefont
  {Basto-Gonzalez}}, \bibinfo {author} {\bibfnamefont {A.}~\bibnamefont
  {Esmaili}}, \ and\ \bibinfo {author} {\bibfnamefont {O.~L.~G.}\ \bibnamefont
  {Peres}},\ }\href {\doibase 10.1016/j.physletb.2012.11.048} {\bibfield
  {journal} {\bibinfo  {journal} {Phys. Lett. B}\ }\textbf {\bibinfo {volume}
  {718}},\ \bibinfo {pages} {1020} (\bibinfo {year} {2013})},\ \Eprint
  {http://arxiv.org/abs/1205.6212} {arXiv:1205.6212 [hep-ph]} \BibitemShut
  {NoStop}%
\bibitem [{\citenamefont {Girardi}\ and\ \citenamefont
  {Meloni}(2014)}]{Girardi:2014gna}%
  \BibitemOpen
  \bibfield  {author} {\bibinfo {author} {\bibfnamefont {I.}~\bibnamefont
  {Girardi}}\ and\ \bibinfo {author} {\bibfnamefont {D.}~\bibnamefont
  {Meloni}},\ }\href {\doibase 10.1103/PhysRevD.90.073011} {\bibfield
  {journal} {\bibinfo  {journal} {Phys. Rev. D}\ }\textbf {\bibinfo {volume}
  {90}},\ \bibinfo {pages} {073011} (\bibinfo {year} {2014})},\ \Eprint
  {http://arxiv.org/abs/1403.5507} {arXiv:1403.5507 [hep-ph]} \BibitemShut
  {NoStop}%
\bibitem [{\citenamefont {Rodejohann}\ and\ \citenamefont
  {Zhang}(2014)}]{Rodejohann:2014eka}%
  \BibitemOpen
  \bibfield  {author} {\bibinfo {author} {\bibfnamefont {W.}~\bibnamefont
  {Rodejohann}}\ and\ \bibinfo {author} {\bibfnamefont {H.}~\bibnamefont
  {Zhang}},\ }\href {\doibase 10.1016/j.physletb.2014.08.035} {\bibfield
  {journal} {\bibinfo  {journal} {Phys. Lett. B}\ }\textbf {\bibinfo {volume}
  {737}},\ \bibinfo {pages} {81} (\bibinfo {year} {2014})},\ \Eprint
  {http://arxiv.org/abs/1407.2739} {arXiv:1407.2739 [hep-ph]} \BibitemShut
  {NoStop}%
\bibitem [{\citenamefont {Berryman}\ \emph {et~al.}(2016)\citenamefont
  {Berryman}, \citenamefont {de~Gouv\^ea}, \citenamefont {Kelly}, \citenamefont
  {Peres},\ and\ \citenamefont {Tabrizi}}]{Berryman:2016szd}%
  \BibitemOpen
  \bibfield  {author} {\bibinfo {author} {\bibfnamefont {J.~M.}\ \bibnamefont
  {Berryman}}, \bibinfo {author} {\bibfnamefont {A.}~\bibnamefont
  {de~Gouv\^ea}}, \bibinfo {author} {\bibfnamefont {K.~J.}\ \bibnamefont
  {Kelly}}, \bibinfo {author} {\bibfnamefont {O.~L.~G.}\ \bibnamefont {Peres}},
  \ and\ \bibinfo {author} {\bibfnamefont {Z.}~\bibnamefont {Tabrizi}},\ }\href
  {\doibase 10.1103/PhysRevD.94.033006} {\bibfield  {journal} {\bibinfo
  {journal} {Phys. Rev. D}\ }\textbf {\bibinfo {volume} {94}},\ \bibinfo
  {pages} {033006} (\bibinfo {year} {2016})},\ \Eprint
  {http://arxiv.org/abs/1603.00018} {arXiv:1603.00018 [hep-ph]} \BibitemShut
  {NoStop}%
\bibitem [{\citenamefont {Carena}\ \emph {et~al.}(2017)\citenamefont {Carena},
  \citenamefont {Li}, \citenamefont {Machado}, \citenamefont {Machado},\ and\
  \citenamefont {Wagner}}]{Carena:2017qhd}%
  \BibitemOpen
  \bibfield  {author} {\bibinfo {author} {\bibfnamefont {M.}~\bibnamefont
  {Carena}}, \bibinfo {author} {\bibfnamefont {Y.-Y.}\ \bibnamefont {Li}},
  \bibinfo {author} {\bibfnamefont {C.~S.}\ \bibnamefont {Machado}}, \bibinfo
  {author} {\bibfnamefont {P.~A.~N.}\ \bibnamefont {Machado}}, \ and\ \bibinfo
  {author} {\bibfnamefont {C.~E.~M.}\ \bibnamefont {Wagner}},\ }\href {\doibase
  10.1103/PhysRevD.96.095014} {\bibfield  {journal} {\bibinfo  {journal} {Phys.
  Rev. D}\ }\textbf {\bibinfo {volume} {96}},\ \bibinfo {pages} {095014}
  (\bibinfo {year} {2017})},\ \Eprint {http://arxiv.org/abs/1708.09548}
  {arXiv:1708.09548 [hep-ph]} \BibitemShut {NoStop}%
\bibitem [{\citenamefont {Stenico}\ \emph {et~al.}(2018)\citenamefont
  {Stenico}, \citenamefont {Forero},\ and\ \citenamefont
  {Peres}}]{Stenico:2018jpl}%
  \BibitemOpen
  \bibfield  {author} {\bibinfo {author} {\bibfnamefont {G.~V.}\ \bibnamefont
  {Stenico}}, \bibinfo {author} {\bibfnamefont {D.~V.}\ \bibnamefont {Forero}},
  \ and\ \bibinfo {author} {\bibfnamefont {O.~L.~G.}\ \bibnamefont {Peres}},\
  }\href {\doibase 10.1007/JHEP11(2018)155} {\bibfield  {journal} {\bibinfo
  {journal} {JHEP}\ }\textbf {\bibinfo {volume} {11}},\ \bibinfo {pages} {155}
  (\bibinfo {year} {2018})},\ \Eprint {http://arxiv.org/abs/1808.05450}
  {arXiv:1808.05450 [hep-ph]} \BibitemShut {NoStop}%
\bibitem [{\citenamefont {Arg\"uelles}\ \emph {et~al.}(2020)\citenamefont
  {Arg\"uelles} \emph {et~al.}}]{Arguelles:2019xgp}%
  \BibitemOpen
  \bibfield  {author} {\bibinfo {author} {\bibfnamefont {C.~A.}\ \bibnamefont
  {Arg\"uelles}} \emph {et~al.},\ }\href {\doibase 10.1088/1361-6633/ab9d12}
  {\bibfield  {journal} {\bibinfo  {journal} {Rept. Prog. Phys.}\ }\textbf
  {\bibinfo {volume} {83}},\ \bibinfo {pages} {124201} (\bibinfo {year}
  {2020})},\ \Eprint {http://arxiv.org/abs/1907.08311} {arXiv:1907.08311
  [hep-ph]} \BibitemShut {NoStop}%
\bibitem [{\citenamefont {Abi}\ \emph {et~al.}(2021)\citenamefont {Abi} \emph
  {et~al.}}]{DUNE:2020fgq}%
  \BibitemOpen
  \bibfield  {author} {\bibinfo {author} {\bibfnamefont {B.}~\bibnamefont
  {Abi}} \emph {et~al.} (\bibinfo {collaboration} {DUNE}),\ }\href {\doibase
  10.1140/epjc/s10052-021-09007-w} {\bibfield  {journal} {\bibinfo  {journal}
  {Eur. Phys. J. C}\ }\textbf {\bibinfo {volume} {81}},\ \bibinfo {pages} {322}
  (\bibinfo {year} {2021})},\ \Eprint {http://arxiv.org/abs/2008.12769}
  {arXiv:2008.12769 [hep-ex]} \BibitemShut {NoStop}%
\bibitem [{\citenamefont {Basto-Gonzalez}\ \emph {et~al.}(2022)\citenamefont
  {Basto-Gonzalez}, \citenamefont {Forero}, \citenamefont {Giunti},
  \citenamefont {Quiroga},\ and\ \citenamefont
  {Ternes}}]{Basto-Gonzalez:2021aus}%
  \BibitemOpen
  \bibfield  {author} {\bibinfo {author} {\bibfnamefont {V.~S.}\ \bibnamefont
  {Basto-Gonzalez}}, \bibinfo {author} {\bibfnamefont {D.~V.}\ \bibnamefont
  {Forero}}, \bibinfo {author} {\bibfnamefont {C.}~\bibnamefont {Giunti}},
  \bibinfo {author} {\bibfnamefont {A.~A.}\ \bibnamefont {Quiroga}}, \ and\
  \bibinfo {author} {\bibfnamefont {C.~A.}\ \bibnamefont {Ternes}},\ }\href
  {\doibase 10.1103/PhysRevD.105.075023} {\bibfield  {journal} {\bibinfo
  {journal} {Phys. Rev. D}\ }\textbf {\bibinfo {volume} {105}},\ \bibinfo
  {pages} {075023} (\bibinfo {year} {2022})},\ \Eprint
  {http://arxiv.org/abs/2112.00379} {arXiv:2112.00379 [hep-ph]} \BibitemShut
  {NoStop}%
\bibitem [{\citenamefont {Arg\"uelles}\ \emph {et~al.}(2022)\citenamefont
  {Arg\"uelles} \emph {et~al.}}]{Arguelles:2022xxa}%
  \BibitemOpen
  \bibfield  {author} {\bibinfo {author} {\bibfnamefont {C.~A.}\ \bibnamefont
  {Arg\"uelles}} \emph {et~al.},\ }in\ \href@noop {} {\emph {\bibinfo
  {booktitle} {{2022 Snowmass Summer Study}}}}\ (\bibinfo {year} {2022})\
  \Eprint {http://arxiv.org/abs/2203.10811} {arXiv:2203.10811 [hep-ph]}
  \BibitemShut {NoStop}%
\bibitem [{\citenamefont {Mention}\ \emph {et~al.}(2011)\citenamefont
  {Mention}, \citenamefont {Fechner}, \citenamefont {Lasserre}, \citenamefont
  {Mueller}, \citenamefont {Lhuillier}, \citenamefont {Cribier},\ and\
  \citenamefont {Letourneau}}]{Mention:2011rk}%
  \BibitemOpen
  \bibfield  {author} {\bibinfo {author} {\bibfnamefont {G.}~\bibnamefont
  {Mention}}, \bibinfo {author} {\bibfnamefont {M.}~\bibnamefont {Fechner}},
  \bibinfo {author} {\bibfnamefont {T.}~\bibnamefont {Lasserre}}, \bibinfo
  {author} {\bibfnamefont {T.~A.}\ \bibnamefont {Mueller}}, \bibinfo {author}
  {\bibfnamefont {D.}~\bibnamefont {Lhuillier}}, \bibinfo {author}
  {\bibfnamefont {M.}~\bibnamefont {Cribier}}, \ and\ \bibinfo {author}
  {\bibfnamefont {A.}~\bibnamefont {Letourneau}},\ }\href {\doibase
  10.1103/PhysRevD.83.073006} {\bibfield  {journal} {\bibinfo  {journal} {Phys.
  Rev. D}\ }\textbf {\bibinfo {volume} {83}},\ \bibinfo {pages} {073006}
  (\bibinfo {year} {2011})},\ \Eprint {http://arxiv.org/abs/1101.2755}
  {arXiv:1101.2755 [hep-ex]} \BibitemShut {NoStop}%
\bibitem [{\citenamefont {Abdurashitov}\ \emph {et~al.}(2006)\citenamefont
  {Abdurashitov} \emph {et~al.}}]{Abdurashitov:2005tb}%
  \BibitemOpen
  \bibfield  {author} {\bibinfo {author} {\bibfnamefont {J.~N.}\ \bibnamefont
  {Abdurashitov}} \emph {et~al.},\ }\href {\doibase 10.1103/PhysRevC.73.045805}
  {\bibfield  {journal} {\bibinfo  {journal} {Phys. Rev. C}\ }\textbf {\bibinfo
  {volume} {73}},\ \bibinfo {pages} {045805} (\bibinfo {year} {2006})},\
  \Eprint {http://arxiv.org/abs/nucl-ex/0512041} {arXiv:nucl-ex/0512041}
  \BibitemShut {NoStop}%
\bibitem [{\citenamefont {Laveder}(2007)}]{Laveder:2007zz}%
  \BibitemOpen
  \bibfield  {author} {\bibinfo {author} {\bibfnamefont {M.}~\bibnamefont
  {Laveder}},\ }\href {\doibase 10.1016/j.nuclphysbps.2007.02.037} {\bibfield
  {journal} {\bibinfo  {journal} {Nucl. Phys. B Proc. Suppl.}\ }\textbf
  {\bibinfo {volume} {168}},\ \bibinfo {pages} {344} (\bibinfo {year}
  {2007})}\BibitemShut {NoStop}%
\bibitem [{\citenamefont {Giunti}\ and\ \citenamefont
  {Laveder}(2007)}]{Giunti:2006bj}%
  \BibitemOpen
  \bibfield  {author} {\bibinfo {author} {\bibfnamefont {C.}~\bibnamefont
  {Giunti}}\ and\ \bibinfo {author} {\bibfnamefont {M.}~\bibnamefont
  {Laveder}},\ }\href {\doibase 10.1142/S0217732307025455} {\bibfield
  {journal} {\bibinfo  {journal} {Mod. Phys. Lett. A}\ }\textbf {\bibinfo
  {volume} {22}},\ \bibinfo {pages} {2499} (\bibinfo {year} {2007})},\ \Eprint
  {http://arxiv.org/abs/hep-ph/0610352} {arXiv:hep-ph/0610352} \BibitemShut
  {NoStop}%
\bibitem [{\citenamefont {Gariazzo}\ \emph {et~al.}(2016)\citenamefont
  {Gariazzo}, \citenamefont {Giunti}, \citenamefont {Laveder}, \citenamefont
  {Li},\ and\ \citenamefont {Zavanin}}]{Gariazzo:2015rra}%
  \BibitemOpen
  \bibfield  {author} {\bibinfo {author} {\bibfnamefont {S.}~\bibnamefont
  {Gariazzo}}, \bibinfo {author} {\bibfnamefont {C.}~\bibnamefont {Giunti}},
  \bibinfo {author} {\bibfnamefont {M.}~\bibnamefont {Laveder}}, \bibinfo
  {author} {\bibfnamefont {Y.~F.}\ \bibnamefont {Li}}, \ and\ \bibinfo {author}
  {\bibfnamefont {E.~M.}\ \bibnamefont {Zavanin}},\ }\href {\doibase
  10.1088/0954-3899/43/3/033001} {\bibfield  {journal} {\bibinfo  {journal} {J.
  Phys. G}\ }\textbf {\bibinfo {volume} {43}},\ \bibinfo {pages} {033001}
  (\bibinfo {year} {2016})},\ \Eprint {http://arxiv.org/abs/1507.08204}
  {arXiv:1507.08204 [hep-ph]} \BibitemShut {NoStop}%
\bibitem [{\citenamefont {Giunti}\ and\ \citenamefont
  {Lasserre}(2019)}]{Giunti:2019aiy}%
  \BibitemOpen
  \bibfield  {author} {\bibinfo {author} {\bibfnamefont {C.}~\bibnamefont
  {Giunti}}\ and\ \bibinfo {author} {\bibfnamefont {T.}~\bibnamefont
  {Lasserre}},\ }\href {\doibase 10.1146/annurev-nucl-101918-023755} {\bibfield
   {journal} {\bibinfo  {journal} {Ann. Rev. Nucl. Part. Sci.}\ }\textbf
  {\bibinfo {volume} {69}},\ \bibinfo {pages} {163} (\bibinfo {year} {2019})},\
  \Eprint {http://arxiv.org/abs/1901.08330} {arXiv:1901.08330 [hep-ph]}
  \BibitemShut {NoStop}%
\bibitem [{\citenamefont {Diaz}\ \emph {et~al.}(2020)\citenamefont {Diaz},
  \citenamefont {Arg\"uelles}, \citenamefont {Collin}, \citenamefont {Conrad},\
  and\ \citenamefont {Shaevitz}}]{Diaz:2019fwt}%
  \BibitemOpen
  \bibfield  {author} {\bibinfo {author} {\bibfnamefont {A.}~\bibnamefont
  {Diaz}}, \bibinfo {author} {\bibfnamefont {C.~A.}\ \bibnamefont
  {Arg\"uelles}}, \bibinfo {author} {\bibfnamefont {G.~H.}\ \bibnamefont
  {Collin}}, \bibinfo {author} {\bibfnamefont {J.~M.}\ \bibnamefont {Conrad}},
  \ and\ \bibinfo {author} {\bibfnamefont {M.~H.}\ \bibnamefont {Shaevitz}},\
  }\href {\doibase 10.1016/j.physrep.2020.08.005} {\bibfield  {journal}
  {\bibinfo  {journal} {Phys. Rept.}\ }\textbf {\bibinfo {volume} {884}},\
  \bibinfo {pages} {1} (\bibinfo {year} {2020})},\ \Eprint
  {http://arxiv.org/abs/1906.00045} {arXiv:1906.00045 [hep-ex]} \BibitemShut
  {NoStop}%
\bibitem [{\citenamefont {B\"oser}\ \emph {et~al.}(2020)\citenamefont
  {B\"oser}, \citenamefont {Buck}, \citenamefont {Giunti}, \citenamefont
  {Lesgourgues}, \citenamefont {Ludhova}, \citenamefont {Mertens},
  \citenamefont {Schukraft},\ and\ \citenamefont {Wurm}}]{Boser:2019rta}%
  \BibitemOpen
  \bibfield  {author} {\bibinfo {author} {\bibfnamefont {S.}~\bibnamefont
  {B\"oser}}, \bibinfo {author} {\bibfnamefont {C.}~\bibnamefont {Buck}},
  \bibinfo {author} {\bibfnamefont {C.}~\bibnamefont {Giunti}}, \bibinfo
  {author} {\bibfnamefont {J.}~\bibnamefont {Lesgourgues}}, \bibinfo {author}
  {\bibfnamefont {L.}~\bibnamefont {Ludhova}}, \bibinfo {author} {\bibfnamefont
  {S.}~\bibnamefont {Mertens}}, \bibinfo {author} {\bibfnamefont
  {A.}~\bibnamefont {Schukraft}}, \ and\ \bibinfo {author} {\bibfnamefont
  {M.}~\bibnamefont {Wurm}},\ }\href {\doibase 10.1016/j.ppnp.2019.103736}
  {\bibfield  {journal} {\bibinfo  {journal} {Prog. Part. Nucl. Phys.}\
  }\textbf {\bibinfo {volume} {111}},\ \bibinfo {pages} {103736} (\bibinfo
  {year} {2020})},\ \Eprint {http://arxiv.org/abs/1906.01739} {arXiv:1906.01739
  [hep-ex]} \BibitemShut {NoStop}%
\bibitem [{\citenamefont {Dasgupta}\ and\ \citenamefont
  {Kopp}(2021)}]{Dasgupta:2021ies}%
  \BibitemOpen
  \bibfield  {author} {\bibinfo {author} {\bibfnamefont {B.}~\bibnamefont
  {Dasgupta}}\ and\ \bibinfo {author} {\bibfnamefont {J.}~\bibnamefont
  {Kopp}},\ }\href {\doibase 10.1016/j.physrep.2021.06.002} {\bibfield
  {journal} {\bibinfo  {journal} {Phys. Rept.}\ }\textbf {\bibinfo {volume}
  {928}},\ \bibinfo {pages} {63} (\bibinfo {year} {2021})},\ \Eprint
  {http://arxiv.org/abs/2106.05913} {arXiv:2106.05913 [hep-ph]} \BibitemShut
  {NoStop}%
\bibitem [{\citenamefont {Barinov}\ \emph {et~al.}(2021)\citenamefont {Barinov}
  \emph {et~al.}}]{Barinov:2021asz}%
  \BibitemOpen
  \bibfield  {author} {\bibinfo {author} {\bibfnamefont {V.~V.}\ \bibnamefont
  {Barinov}} \emph {et~al.},\ }\href@noop {} {\  (\bibinfo {year} {2021})},\
  \Eprint {http://arxiv.org/abs/2109.11482} {arXiv:2109.11482 [nucl-ex]}
  \BibitemShut {NoStop}%
\bibitem [{\citenamefont {Barinov}\ \emph {et~al.}(2022)\citenamefont {Barinov}
  \emph {et~al.}}]{Barinov:2022wfh}%
  \BibitemOpen
  \bibfield  {author} {\bibinfo {author} {\bibfnamefont {V.~V.}\ \bibnamefont
  {Barinov}} \emph {et~al.},\ }\href@noop {} {\  (\bibinfo {year} {2022})},\
  \Eprint {http://arxiv.org/abs/2201.07364} {arXiv:2201.07364 [nucl-ex]}
  \BibitemShut {NoStop}%
\bibitem [{\citenamefont {Gariazzo}\ \emph {et~al.}(2022)\citenamefont
  {Gariazzo} \emph {et~al.}}]{Gariazzo:2022ahe}%
  \BibitemOpen
  \bibfield  {author} {\bibinfo {author} {\bibfnamefont {S.}~\bibnamefont
  {Gariazzo}} \emph {et~al.},\ }\href@noop {} {\  (\bibinfo {year} {2022})},\
  \Eprint {http://arxiv.org/abs/2205.02195} {arXiv:2205.02195 [hep-ph]}
  \BibitemShut {NoStop}%
\bibitem [{\citenamefont {Giunti}\ \emph {et~al.}(2022)\citenamefont {Giunti},
  \citenamefont {Li}, \citenamefont {Ternes},\ and\ \citenamefont
  {Xin}}]{Giunti:2021kab}%
  \BibitemOpen
  \bibfield  {author} {\bibinfo {author} {\bibfnamefont {C.}~\bibnamefont
  {Giunti}}, \bibinfo {author} {\bibfnamefont {Y.~F.}\ \bibnamefont {Li}},
  \bibinfo {author} {\bibfnamefont {C.~A.}\ \bibnamefont {Ternes}}, \ and\
  \bibinfo {author} {\bibfnamefont {Z.}~\bibnamefont {Xin}},\ }\href {\doibase
  10.1016/j.physletb.2022.137054} {\bibfield  {journal} {\bibinfo  {journal}
  {Phys. Lett. B}\ }\textbf {\bibinfo {volume} {829}},\ \bibinfo {pages}
  {137054} (\bibinfo {year} {2022})},\ \Eprint
  {http://arxiv.org/abs/2110.06820} {arXiv:2110.06820 [hep-ph]} \BibitemShut
  {NoStop}%
\bibitem [{\citenamefont {Berryman}\ and\ \citenamefont
  {Huber}(2021)}]{Berryman:2020agd}%
  \BibitemOpen
  \bibfield  {author} {\bibinfo {author} {\bibfnamefont {J.~M.}\ \bibnamefont
  {Berryman}}\ and\ \bibinfo {author} {\bibfnamefont {P.}~\bibnamefont
  {Huber}},\ }\href {\doibase 10.1007/JHEP01(2021)167} {\bibfield  {journal}
  {\bibinfo  {journal} {JHEP}\ }\textbf {\bibinfo {volume} {01}},\ \bibinfo
  {pages} {167} (\bibinfo {year} {2021})},\ \Eprint
  {http://arxiv.org/abs/2005.01756} {arXiv:2005.01756 [hep-ph]} \BibitemShut
  {NoStop}%
\bibitem [{\citenamefont {Mueller}\ \emph {et~al.}(2011)\citenamefont {Mueller}
  \emph {et~al.}}]{Mueller:2011nm}%
  \BibitemOpen
  \bibfield  {author} {\bibinfo {author} {\bibfnamefont {T.~A.}\ \bibnamefont
  {Mueller}} \emph {et~al.},\ }\href@noop {} {\bibfield  {journal} {\bibinfo
  {journal} {Phys. Rev.}\ }\textbf {\bibinfo {volume} {C83}},\ \bibinfo {pages}
  {054615} (\bibinfo {year} {2011})},\ \Eprint
  {http://arxiv.org/abs/arXiv:1101.2663} {arXiv:1101.2663 [hep-ex]}
  \BibitemShut {NoStop}%
\bibitem [{\citenamefont {Huber}(2011)}]{Huber:2011wv}%
  \BibitemOpen
  \bibfield  {author} {\bibinfo {author} {\bibfnamefont {P.}~\bibnamefont
  {Huber}},\ }\href@noop {} {\bibfield  {journal} {\bibinfo  {journal} {Phys.
  Rev.}\ }\textbf {\bibinfo {volume} {C84}},\ \bibinfo {pages} {024617}
  (\bibinfo {year} {2011})},\ \Eprint {http://arxiv.org/abs/arXiv:1106.0687}
  {arXiv:1106.0687 [hep-ph]} \BibitemShut {NoStop}%
\bibitem [{\citenamefont {Estienne}\ \emph {et~al.}(2019)\citenamefont
  {Estienne}, \citenamefont {Fallot} \emph {et~al.}}]{Estienne:2019ujo}%
  \BibitemOpen
  \bibfield  {author} {\bibinfo {author} {\bibfnamefont {M.}~\bibnamefont
  {Estienne}}, \bibinfo {author} {\bibfnamefont {M.}~\bibnamefont {Fallot}},
  \emph {et~al.},\ }\href {\doibase 10.1103/PhysRevLett.123.022502} {\bibfield
  {journal} {\bibinfo  {journal} {Phys. Rev. Lett.}\ }\textbf {\bibinfo
  {volume} {123}},\ \bibinfo {pages} {022502} (\bibinfo {year} {2019})},\
  \Eprint {http://arxiv.org/abs/arXiv:1904.09358} {arXiv:1904.09358 [nucl-ex]}
  \BibitemShut {NoStop}%
\bibitem [{\citenamefont {Hayen}\ \emph {et~al.}(2019)\citenamefont {Hayen},
  \citenamefont {Kostensalo}, \citenamefont {Severijns},\ and\ \citenamefont
  {Suhonen}}]{Hayen:2019eop}%
  \BibitemOpen
  \bibfield  {author} {\bibinfo {author} {\bibfnamefont {L.}~\bibnamefont
  {Hayen}}, \bibinfo {author} {\bibfnamefont {J.}~\bibnamefont {Kostensalo}},
  \bibinfo {author} {\bibfnamefont {N.}~\bibnamefont {Severijns}}, \ and\
  \bibinfo {author} {\bibfnamefont {J.}~\bibnamefont {Suhonen}},\ }\href@noop
  {} {\bibfield  {journal} {\bibinfo  {journal} {Phys.Rev.}\ }\textbf {\bibinfo
  {volume} {C100}},\ \bibinfo {pages} {054323} (\bibinfo {year} {2019})},\
  \Eprint {http://arxiv.org/abs/arXiv:1908.08302} {arXiv:1908.08302 [nucl-th]}
  \BibitemShut {NoStop}%
\bibitem [{\citenamefont {Kopeikin}\ \emph {et~al.}(2021)\citenamefont
  {Kopeikin}, \citenamefont {Skorokhvatov},\ and\ \citenamefont
  {Titov}}]{Kopeikin:2021ugh}%
  \BibitemOpen
  \bibfield  {author} {\bibinfo {author} {\bibfnamefont {V.}~\bibnamefont
  {Kopeikin}}, \bibinfo {author} {\bibfnamefont {M.}~\bibnamefont
  {Skorokhvatov}}, \ and\ \bibinfo {author} {\bibfnamefont {O.}~\bibnamefont
  {Titov}},\ }\href {\doibase 10.1103/PhysRevD.104.L071301} {\bibfield
  {journal} {\bibinfo  {journal} {Phys. Rev. D}\ }\textbf {\bibinfo {volume}
  {104}},\ \bibinfo {pages} {L071301} (\bibinfo {year} {2021})},\ \Eprint
  {http://arxiv.org/abs/2103.01684} {arXiv:2103.01684 [nucl-ex]} \BibitemShut
  {NoStop}%
\bibitem [{\citenamefont {Anselmann}\ \emph {et~al.}(1995)\citenamefont
  {Anselmann} \emph {et~al.}}]{GALLEX:1994rym}%
  \BibitemOpen
  \bibfield  {author} {\bibinfo {author} {\bibfnamefont {P.}~\bibnamefont
  {Anselmann}} \emph {et~al.} (\bibinfo {collaboration} {GALLEX}),\ }\href
  {\doibase 10.1016/0370-2693(94)01586-2} {\bibfield  {journal} {\bibinfo
  {journal} {Phys. Lett. B}\ }\textbf {\bibinfo {volume} {342}},\ \bibinfo
  {pages} {440} (\bibinfo {year} {1995})}\BibitemShut {NoStop}%
\bibitem [{\citenamefont {Hampel}\ \emph {et~al.}(1998)\citenamefont {Hampel}
  \emph {et~al.}}]{GALLEX:1997lja}%
  \BibitemOpen
  \bibfield  {author} {\bibinfo {author} {\bibfnamefont {W.}~\bibnamefont
  {Hampel}} \emph {et~al.} (\bibinfo {collaboration} {GALLEX}),\ }\href
  {\doibase 10.1016/S0370-2693(97)01562-1} {\bibfield  {journal} {\bibinfo
  {journal} {Phys. Lett. B}\ }\textbf {\bibinfo {volume} {420}},\ \bibinfo
  {pages} {114} (\bibinfo {year} {1998})}\BibitemShut {NoStop}%
\bibitem [{\citenamefont {Kaether}\ \emph {et~al.}(2010)\citenamefont
  {Kaether}, \citenamefont {Hampel}, \citenamefont {Heusser}, \citenamefont
  {Kiko},\ and\ \citenamefont {Kirsten}}]{Kaether:2010ag}%
  \BibitemOpen
  \bibfield  {author} {\bibinfo {author} {\bibfnamefont {F.}~\bibnamefont
  {Kaether}}, \bibinfo {author} {\bibfnamefont {W.}~\bibnamefont {Hampel}},
  \bibinfo {author} {\bibfnamefont {G.}~\bibnamefont {Heusser}}, \bibinfo
  {author} {\bibfnamefont {J.}~\bibnamefont {Kiko}}, \ and\ \bibinfo {author}
  {\bibfnamefont {T.}~\bibnamefont {Kirsten}},\ }\href {\doibase
  10.1016/j.physletb.2010.01.030} {\bibfield  {journal} {\bibinfo  {journal}
  {Phys. Lett. B}\ }\textbf {\bibinfo {volume} {685}},\ \bibinfo {pages} {47}
  (\bibinfo {year} {2010})},\ \Eprint {http://arxiv.org/abs/1001.2731}
  {arXiv:1001.2731 [hep-ex]} \BibitemShut {NoStop}%
\bibitem [{\citenamefont {Abdurashitov}\ \emph {et~al.}(1996)\citenamefont
  {Abdurashitov} \emph {et~al.}}]{Abdurashitov:1996dp}%
  \BibitemOpen
  \bibfield  {author} {\bibinfo {author} {\bibfnamefont {D.~N.}\ \bibnamefont
  {Abdurashitov}} \emph {et~al.},\ }\href {\doibase
  10.1103/PhysRevLett.77.4708} {\bibfield  {journal} {\bibinfo  {journal}
  {Phys. Rev. Lett.}\ }\textbf {\bibinfo {volume} {77}},\ \bibinfo {pages}
  {4708} (\bibinfo {year} {1996})}\BibitemShut {NoStop}%
\bibitem [{\citenamefont {Abdurashitov}\ \emph {et~al.}(1999)\citenamefont
  {Abdurashitov} \emph {et~al.}}]{SAGE:1998fvr}%
  \BibitemOpen
  \bibfield  {author} {\bibinfo {author} {\bibfnamefont {J.~N.}\ \bibnamefont
  {Abdurashitov}} \emph {et~al.} (\bibinfo {collaboration} {SAGE}),\ }\href
  {\doibase 10.1103/PhysRevC.59.2246} {\bibfield  {journal} {\bibinfo
  {journal} {Phys. Rev. C}\ }\textbf {\bibinfo {volume} {59}},\ \bibinfo
  {pages} {2246} (\bibinfo {year} {1999})},\ \Eprint
  {http://arxiv.org/abs/hep-ph/9803418} {arXiv:hep-ph/9803418} \BibitemShut
  {NoStop}%
\bibitem [{\citenamefont {Abdurashitov}\ \emph {et~al.}(2009)\citenamefont
  {Abdurashitov} \emph {et~al.}}]{SAGE:2009eeu}%
  \BibitemOpen
  \bibfield  {author} {\bibinfo {author} {\bibfnamefont {J.~N.}\ \bibnamefont
  {Abdurashitov}} \emph {et~al.} (\bibinfo {collaboration} {SAGE}),\ }\href
  {\doibase 10.1103/PhysRevC.80.015807} {\bibfield  {journal} {\bibinfo
  {journal} {Phys. Rev. C}\ }\textbf {\bibinfo {volume} {80}},\ \bibinfo
  {pages} {015807} (\bibinfo {year} {2009})},\ \Eprint
  {http://arxiv.org/abs/0901.2200} {arXiv:0901.2200 [nucl-ex]} \BibitemShut
  {NoStop}%
\bibitem [{\citenamefont {Acero}\ \emph {et~al.}(2008)\citenamefont {Acero},
  \citenamefont {Giunti},\ and\ \citenamefont {Laveder}}]{Acero:2007su}%
  \BibitemOpen
  \bibfield  {author} {\bibinfo {author} {\bibfnamefont {M.~A.}\ \bibnamefont
  {Acero}}, \bibinfo {author} {\bibfnamefont {C.}~\bibnamefont {Giunti}}, \
  and\ \bibinfo {author} {\bibfnamefont {M.}~\bibnamefont {Laveder}},\ }\href
  {\doibase 10.1103/PhysRevD.78.073009} {\bibfield  {journal} {\bibinfo
  {journal} {Phys. Rev. D}\ }\textbf {\bibinfo {volume} {78}},\ \bibinfo
  {pages} {073009} (\bibinfo {year} {2008})},\ \Eprint
  {http://arxiv.org/abs/0711.4222} {arXiv:0711.4222 [hep-ph]} \BibitemShut
  {NoStop}%
\bibitem [{\citenamefont {Bahcall}(1997)}]{Bahcall:1997eg}%
  \BibitemOpen
  \bibfield  {author} {\bibinfo {author} {\bibfnamefont {J.~N.}\ \bibnamefont
  {Bahcall}},\ }\href {\doibase 10.1103/PhysRevC.56.3391} {\bibfield  {journal}
  {\bibinfo  {journal} {Phys. Rev. C}\ }\textbf {\bibinfo {volume} {56}},\
  \bibinfo {pages} {3391} (\bibinfo {year} {1997})},\ \Eprint
  {http://arxiv.org/abs/hep-ph/9710491} {arXiv:hep-ph/9710491} \BibitemShut
  {NoStop}%
\bibitem [{\citenamefont {Berryman}\ \emph {et~al.}(2022)\citenamefont
  {Berryman}, \citenamefont {Coloma}, \citenamefont {Huber}, \citenamefont
  {Schwetz},\ and\ \citenamefont {Zhou}}]{Berryman:2021yan}%
  \BibitemOpen
  \bibfield  {author} {\bibinfo {author} {\bibfnamefont {J.~M.}\ \bibnamefont
  {Berryman}}, \bibinfo {author} {\bibfnamefont {P.}~\bibnamefont {Coloma}},
  \bibinfo {author} {\bibfnamefont {P.}~\bibnamefont {Huber}}, \bibinfo
  {author} {\bibfnamefont {T.}~\bibnamefont {Schwetz}}, \ and\ \bibinfo
  {author} {\bibfnamefont {A.}~\bibnamefont {Zhou}},\ }\href {\doibase
  10.1007/JHEP02(2022)055} {\bibfield  {journal} {\bibinfo  {journal} {JHEP}\
  }\textbf {\bibinfo {volume} {02}},\ \bibinfo {pages} {055} (\bibinfo {year}
  {2022})},\ \Eprint {http://arxiv.org/abs/2111.12530} {arXiv:2111.12530
  [hep-ph]} \BibitemShut {NoStop}%
\bibitem [{\citenamefont {Adamson}\ \emph {et~al.}(2016)\citenamefont {Adamson}
  \emph {et~al.}}]{MINOS:2016vvv}%
  \BibitemOpen
  \bibfield  {author} {\bibinfo {author} {\bibfnamefont {P.}~\bibnamefont
  {Adamson}} \emph {et~al.} (\bibinfo {collaboration} {MINOS}),\ }\href
  {\doibase 10.1103/PhysRevD.94.111101} {\bibfield  {journal} {\bibinfo
  {journal} {Phys. Rev. D}\ }\textbf {\bibinfo {volume} {94}},\ \bibinfo
  {pages} {111101} (\bibinfo {year} {2016})},\ \Eprint
  {http://arxiv.org/abs/1608.06964} {arXiv:1608.06964 [hep-ex]} \BibitemShut
  {NoStop}%
\bibitem [{\citenamefont {Adamson}\ \emph {et~al.}(2019)\citenamefont {Adamson}
  \emph {et~al.}}]{MINOS:2017cae}%
  \BibitemOpen
  \bibfield  {author} {\bibinfo {author} {\bibfnamefont {P.}~\bibnamefont
  {Adamson}} \emph {et~al.} (\bibinfo {collaboration} {MINOS+}),\ }\href
  {\doibase 10.1103/PhysRevLett.122.091803} {\bibfield  {journal} {\bibinfo
  {journal} {Phys. Rev. Lett.}\ }\textbf {\bibinfo {volume} {122}},\ \bibinfo
  {pages} {091803} (\bibinfo {year} {2019})},\ \Eprint
  {http://arxiv.org/abs/1710.06488} {arXiv:1710.06488 [hep-ex]} \BibitemShut
  {NoStop}%
\bibitem [{\citenamefont {Adey}\ \emph {et~al.}(2018)\citenamefont {Adey} \emph
  {et~al.}}]{DayaBay:2018yms}%
  \BibitemOpen
  \bibfield  {author} {\bibinfo {author} {\bibfnamefont {D.}~\bibnamefont
  {Adey}} \emph {et~al.} (\bibinfo {collaboration} {Daya Bay}),\ }\href
  {\doibase 10.1103/PhysRevLett.121.241805} {\bibfield  {journal} {\bibinfo
  {journal} {Phys. Rev. Lett.}\ }\textbf {\bibinfo {volume} {121}},\ \bibinfo
  {pages} {241805} (\bibinfo {year} {2018})},\ \Eprint
  {http://arxiv.org/abs/1809.02261} {arXiv:1809.02261 [hep-ex]} \BibitemShut
  {NoStop}%
\bibitem [{\citenamefont {An}\ \emph {et~al.}(2017{\natexlab{a}})\citenamefont
  {An} \emph {et~al.}}]{DayaBay:2016ssb}%
  \BibitemOpen
  \bibfield  {author} {\bibinfo {author} {\bibfnamefont {F.~P.}\ \bibnamefont
  {An}} \emph {et~al.} (\bibinfo {collaboration} {Daya Bay}),\ }\href {\doibase
  10.1088/1674-1137/41/1/013002} {\bibfield  {journal} {\bibinfo  {journal}
  {Chin. Phys. C}\ }\textbf {\bibinfo {volume} {41}},\ \bibinfo {pages}
  {013002} (\bibinfo {year} {2017}{\natexlab{a}})},\ \Eprint
  {http://arxiv.org/abs/1607.05378} {arXiv:1607.05378 [hep-ex]} \BibitemShut
  {NoStop}%
\bibitem [{\citenamefont {An}\ \emph {et~al.}(2017{\natexlab{b}})\citenamefont
  {An} \emph {et~al.}}]{DayaBay:2016ggj}%
  \BibitemOpen
  \bibfield  {author} {\bibinfo {author} {\bibfnamefont {F.~P.}\ \bibnamefont
  {An}} \emph {et~al.} (\bibinfo {collaboration} {Daya Bay}),\ }\href {\doibase
  10.1103/PhysRevD.95.072006} {\bibfield  {journal} {\bibinfo  {journal} {Phys.
  Rev. D}\ }\textbf {\bibinfo {volume} {95}},\ \bibinfo {pages} {072006}
  (\bibinfo {year} {2017}{\natexlab{b}})},\ \Eprint
  {http://arxiv.org/abs/1610.04802} {arXiv:1610.04802 [hep-ex]} \BibitemShut
  {NoStop}%
\bibitem [{\citenamefont {Huber}\ \emph {et~al.}(2005)\citenamefont {Huber},
  \citenamefont {Lindner},\ and\ \citenamefont {Winter}}]{Huber:2004ka}%
  \BibitemOpen
  \bibfield  {author} {\bibinfo {author} {\bibfnamefont {P.}~\bibnamefont
  {Huber}}, \bibinfo {author} {\bibfnamefont {M.}~\bibnamefont {Lindner}}, \
  and\ \bibinfo {author} {\bibfnamefont {W.}~\bibnamefont {Winter}},\ }\href
  {\doibase 10.1016/j.cpc.2005.01.003} {\bibfield  {journal} {\bibinfo
  {journal} {Comput. Phys. Commun.}\ }\textbf {\bibinfo {volume} {167}},\
  \bibinfo {pages} {195} (\bibinfo {year} {2005})},\ \Eprint
  {http://arxiv.org/abs/hep-ph/0407333} {arXiv:hep-ph/0407333} \BibitemShut
  {NoStop}%
\bibitem [{\citenamefont {Huber}\ \emph {et~al.}(2007)\citenamefont {Huber},
  \citenamefont {Kopp}, \citenamefont {Lindner}, \citenamefont {Rolinec},\ and\
  \citenamefont {Winter}}]{Huber:2007ji}%
  \BibitemOpen
  \bibfield  {author} {\bibinfo {author} {\bibfnamefont {P.}~\bibnamefont
  {Huber}}, \bibinfo {author} {\bibfnamefont {J.}~\bibnamefont {Kopp}},
  \bibinfo {author} {\bibfnamefont {M.}~\bibnamefont {Lindner}}, \bibinfo
  {author} {\bibfnamefont {M.}~\bibnamefont {Rolinec}}, \ and\ \bibinfo
  {author} {\bibfnamefont {W.}~\bibnamefont {Winter}},\ }\href {\doibase
  10.1016/j.cpc.2007.05.004} {\bibfield  {journal} {\bibinfo  {journal}
  {Comput. Phys. Commun.}\ }\textbf {\bibinfo {volume} {177}},\ \bibinfo
  {pages} {432} (\bibinfo {year} {2007})},\ \Eprint
  {http://arxiv.org/abs/hep-ph/0701187} {arXiv:hep-ph/0701187} \BibitemShut
  {NoStop}%
\bibitem [{\citenamefont {De~Rijck}\ and\ \citenamefont
  {Huang}(2017)}]{DeRijck:2017wuy}%
  \BibitemOpen
  \bibfield  {author} {\bibinfo {author} {\bibfnamefont {S.}~\bibnamefont
  {De~Rijck}}\ and\ \bibinfo {author} {\bibfnamefont {J.}~\bibnamefont
  {Huang}},\ }\href {\doibase 10.1088/1742-6596/888/1/012162} {\bibfield
  {journal} {\bibinfo  {journal} {J. Phys. Conf. Ser.}\ }\textbf {\bibinfo
  {volume} {888}},\ \bibinfo {pages} {012162} (\bibinfo {year}
  {2017})}\BibitemShut {NoStop}%
\bibitem [{\citenamefont {Aker}\ \emph {et~al.}(2019)\citenamefont {Aker} \emph
  {et~al.}}]{KATRIN:2019yun}%
  \BibitemOpen
  \bibfield  {author} {\bibinfo {author} {\bibfnamefont {M.}~\bibnamefont
  {Aker}} \emph {et~al.} (\bibinfo {collaboration} {KATRIN}),\ }\href {\doibase
  10.1103/PhysRevLett.123.221802} {\bibfield  {journal} {\bibinfo  {journal}
  {Phys. Rev. Lett.}\ }\textbf {\bibinfo {volume} {123}},\ \bibinfo {pages}
  {221802} (\bibinfo {year} {2019})},\ \Eprint
  {http://arxiv.org/abs/1909.06048} {arXiv:1909.06048 [hep-ex]} \BibitemShut
  {NoStop}%
\bibitem [{\citenamefont {Aker}\ \emph
  {et~al.}(2022{\natexlab{a}})\citenamefont {Aker} \emph
  {et~al.}}]{KATRIN:2021uub}%
  \BibitemOpen
  \bibfield  {author} {\bibinfo {author} {\bibfnamefont {M.}~\bibnamefont
  {Aker}} \emph {et~al.} (\bibinfo {collaboration} {KATRIN}),\ }\href {\doibase
  10.1038/s41567-021-01463-1} {\bibfield  {journal} {\bibinfo  {journal}
  {Nature Phys.}\ }\textbf {\bibinfo {volume} {18}},\ \bibinfo {pages} {160}
  (\bibinfo {year} {2022}{\natexlab{a}})},\ \Eprint
  {http://arxiv.org/abs/2105.08533} {arXiv:2105.08533 [hep-ex]} \BibitemShut
  {NoStop}%
\bibitem [{\citenamefont {Giunti}\ \emph {et~al.}(2020)\citenamefont {Giunti},
  \citenamefont {Li},\ and\ \citenamefont {Zhang}}]{Giunti:2019fcj}%
  \BibitemOpen
  \bibfield  {author} {\bibinfo {author} {\bibfnamefont {C.}~\bibnamefont
  {Giunti}}, \bibinfo {author} {\bibfnamefont {Y.~F.}\ \bibnamefont {Li}}, \
  and\ \bibinfo {author} {\bibfnamefont {Y.~Y.}\ \bibnamefont {Zhang}},\ }\href
  {\doibase 10.1007/JHEP05(2020)061} {\bibfield  {journal} {\bibinfo  {journal}
  {JHEP}\ }\textbf {\bibinfo {volume} {05}},\ \bibinfo {pages} {061} (\bibinfo
  {year} {2020})},\ \Eprint {http://arxiv.org/abs/1912.12956} {arXiv:1912.12956
  [hep-ph]} \BibitemShut {NoStop}%
\bibitem [{\citenamefont {Aker}\ \emph {et~al.}(2021)\citenamefont {Aker} \emph
  {et~al.}}]{KATRIN:2020dpx}%
  \BibitemOpen
  \bibfield  {author} {\bibinfo {author} {\bibfnamefont {M.}~\bibnamefont
  {Aker}} \emph {et~al.} (\bibinfo {collaboration} {KATRIN}),\ }\href {\doibase
  10.1103/PhysRevLett.126.091803} {\bibfield  {journal} {\bibinfo  {journal}
  {Phys. Rev. Lett.}\ }\textbf {\bibinfo {volume} {126}},\ \bibinfo {pages}
  {091803} (\bibinfo {year} {2021})},\ \Eprint
  {http://arxiv.org/abs/2011.05087} {arXiv:2011.05087 [hep-ex]} \BibitemShut
  {NoStop}%
\bibitem [{\citenamefont {Aker}\ \emph
  {et~al.}(2022{\natexlab{b}})\citenamefont {Aker} \emph
  {et~al.}}]{KATRIN:2022ith}%
  \BibitemOpen
  \bibfield  {author} {\bibinfo {author} {\bibfnamefont {M.}~\bibnamefont
  {Aker}} \emph {et~al.} (\bibinfo {collaboration} {KATRIN}),\ }\href {\doibase
  10.1103/PhysRevD.105.072004} {\bibfield  {journal} {\bibinfo  {journal}
  {Phys. Rev. D}\ }\textbf {\bibinfo {volume} {105}},\ \bibinfo {pages}
  {072004} (\bibinfo {year} {2022}{\natexlab{b}})},\ \Eprint
  {http://arxiv.org/abs/2201.11593} {arXiv:2201.11593 [hep-ex]} \BibitemShut
  {NoStop}%
\bibitem [{\citenamefont {Myers}\ \emph {et~al.}(2015)\citenamefont {Myers},
  \citenamefont {Wagner}, \citenamefont {Kracke},\ and\ \citenamefont
  {Wesson}}]{Myers:2015lca}%
  \BibitemOpen
  \bibfield  {author} {\bibinfo {author} {\bibfnamefont {E.~G.}\ \bibnamefont
  {Myers}}, \bibinfo {author} {\bibfnamefont {A.}~\bibnamefont {Wagner}},
  \bibinfo {author} {\bibfnamefont {H.}~\bibnamefont {Kracke}}, \ and\ \bibinfo
  {author} {\bibfnamefont {B.~A.}\ \bibnamefont {Wesson}},\ }\href {\doibase
  10.1103/PhysRevLett.114.013003} {\bibfield  {journal} {\bibinfo  {journal}
  {Phys. Rev. Lett.}\ }\textbf {\bibinfo {volume} {114}},\ \bibinfo {pages}
  {013003} (\bibinfo {year} {2015})}\BibitemShut {NoStop}%
\bibitem [{\citenamefont {Di~Iura}\ \emph {et~al.}(2015)\citenamefont
  {Di~Iura}, \citenamefont {Girardi},\ and\ \citenamefont
  {Meloni}}]{DiIura:2014csa}%
  \BibitemOpen
  \bibfield  {author} {\bibinfo {author} {\bibfnamefont {A.}~\bibnamefont
  {Di~Iura}}, \bibinfo {author} {\bibfnamefont {I.}~\bibnamefont {Girardi}}, \
  and\ \bibinfo {author} {\bibfnamefont {D.}~\bibnamefont {Meloni}},\ }\href
  {\doibase 10.1088/0954-3899/42/6/065003} {\bibfield  {journal} {\bibinfo
  {journal} {J. Phys. G}\ }\textbf {\bibinfo {volume} {42}},\ \bibinfo {pages}
  {065003} (\bibinfo {year} {2015})},\ \Eprint {http://arxiv.org/abs/1411.5330}
  {arXiv:1411.5330 [hep-ph]} \BibitemShut {NoStop}%
\bibitem [{\citenamefont {Long}\ \emph {et~al.}(1999)\citenamefont {Long},
  \citenamefont {Chan},\ and\ \citenamefont {Price}}]{Long:1998dk}%
  \BibitemOpen
  \bibfield  {author} {\bibinfo {author} {\bibfnamefont {J.~C.}\ \bibnamefont
  {Long}}, \bibinfo {author} {\bibfnamefont {H.~W.}\ \bibnamefont {Chan}}, \
  and\ \bibinfo {author} {\bibfnamefont {J.~C.}\ \bibnamefont {Price}},\ }\href
  {\doibase 10.1016/S0550-3213(98)00711-1} {\bibfield  {journal} {\bibinfo
  {journal} {Nucl. Phys. B}\ }\textbf {\bibinfo {volume} {539}},\ \bibinfo
  {pages} {23} (\bibinfo {year} {1999})},\ \Eprint
  {http://arxiv.org/abs/hep-ph/9805217} {arXiv:hep-ph/9805217} \BibitemShut
  {NoStop}%
\bibitem [{\citenamefont {Krause}\ and\ \citenamefont
  {Fischbach}(2001)}]{Krause:1999ry}%
  \BibitemOpen
  \bibfield  {author} {\bibinfo {author} {\bibfnamefont {D.~E.}\ \bibnamefont
  {Krause}}\ and\ \bibinfo {author} {\bibfnamefont {E.}~\bibnamefont
  {Fischbach}},\ }\href {\doibase 10.1007/3-540-40988-2_14} {\bibfield
  {journal} {\bibinfo  {journal} {Lect. Notes Phys.}\ }\textbf {\bibinfo
  {volume} {562}},\ \bibinfo {pages} {292} (\bibinfo {year} {2001})},\ \Eprint
  {http://arxiv.org/abs/hep-ph/9912276} {arXiv:hep-ph/9912276} \BibitemShut
  {NoStop}%
\bibitem [{\citenamefont {Fischbach}\ \emph {et~al.}(2001)\citenamefont
  {Fischbach}, \citenamefont {Krause}, \citenamefont {Mostepanenko},\ and\
  \citenamefont {Novello}}]{Fischbach:2001ry}%
  \BibitemOpen
  \bibfield  {author} {\bibinfo {author} {\bibfnamefont {E.}~\bibnamefont
  {Fischbach}}, \bibinfo {author} {\bibfnamefont {D.~E.}\ \bibnamefont
  {Krause}}, \bibinfo {author} {\bibfnamefont {V.~M.}\ \bibnamefont
  {Mostepanenko}}, \ and\ \bibinfo {author} {\bibfnamefont {M.}~\bibnamefont
  {Novello}},\ }\href {\doibase 10.1103/PhysRevD.64.075010} {\bibfield
  {journal} {\bibinfo  {journal} {Phys. Rev. D}\ }\textbf {\bibinfo {volume}
  {64}},\ \bibinfo {pages} {075010} (\bibinfo {year} {2001})},\ \Eprint
  {http://arxiv.org/abs/hep-ph/0106331} {arXiv:hep-ph/0106331} \BibitemShut
  {NoStop}%
\bibitem [{\citenamefont {Adelberger}(2002)}]{Adelberger:2002ic}%
  \BibitemOpen
  \bibfield  {author} {\bibinfo {author} {\bibfnamefont {E.~G.}\ \bibnamefont
  {Adelberger}} (\bibinfo {collaboration} {EOT-WASH Group}),\ }in\ \href
  {\doibase 10.1142/9789812778123_0002} {\emph {\bibinfo {booktitle} {{2nd
  Meeting on CPT and Lorentz Symmetry}}}}\ (\bibinfo {year} {2002})\ pp.\
  \bibinfo {pages} {9--15},\ \Eprint {http://arxiv.org/abs/hep-ex/0202008}
  {arXiv:hep-ex/0202008} \BibitemShut {NoStop}%
\bibitem [{\citenamefont {Rizzo}(1999)}]{Rizzo:1998fm}%
  \BibitemOpen
  \bibfield  {author} {\bibinfo {author} {\bibfnamefont {T.~G.}\ \bibnamefont
  {Rizzo}},\ }\href {\doibase 10.1103/PhysRevD.59.115010} {\bibfield  {journal}
  {\bibinfo  {journal} {Phys. Rev. D}\ }\textbf {\bibinfo {volume} {59}},\
  \bibinfo {pages} {115010} (\bibinfo {year} {1999})},\ \Eprint
  {http://arxiv.org/abs/hep-ph/9901209} {arXiv:hep-ph/9901209} \BibitemShut
  {NoStop}%
\bibitem [{\citenamefont {Hewett}(1999)}]{Hewett:1998sn}%
  \BibitemOpen
  \bibfield  {author} {\bibinfo {author} {\bibfnamefont {J.~L.}\ \bibnamefont
  {Hewett}},\ }\href {\doibase 10.1103/PhysRevLett.82.4765} {\bibfield
  {journal} {\bibinfo  {journal} {Phys. Rev. Lett.}\ }\textbf {\bibinfo
  {volume} {82}},\ \bibinfo {pages} {4765} (\bibinfo {year} {1999})},\ \Eprint
  {http://arxiv.org/abs/hep-ph/9811356} {arXiv:hep-ph/9811356} \BibitemShut
  {NoStop}%
\bibitem [{\citenamefont {Abbott}\ \emph {et~al.}(2001)\citenamefont {Abbott}
  \emph {et~al.}}]{D0:2000cve}%
  \BibitemOpen
  \bibfield  {author} {\bibinfo {author} {\bibfnamefont {B.}~\bibnamefont
  {Abbott}} \emph {et~al.} (\bibinfo {collaboration} {D0}),\ }\href {\doibase
  10.1103/PhysRevLett.86.1156} {\bibfield  {journal} {\bibinfo  {journal}
  {Phys. Rev. Lett.}\ }\textbf {\bibinfo {volume} {86}},\ \bibinfo {pages}
  {1156} (\bibinfo {year} {2001})},\ \Eprint
  {http://arxiv.org/abs/hep-ex/0008065} {arXiv:hep-ex/0008065} \BibitemShut
  {NoStop}%
\bibitem [{\citenamefont {Abreu}\ \emph {et~al.}(2000)\citenamefont {Abreu}
  \emph {et~al.}}]{DELPHI:2000ztm}%
  \BibitemOpen
  \bibfield  {author} {\bibinfo {author} {\bibfnamefont {P.}~\bibnamefont
  {Abreu}} \emph {et~al.} (\bibinfo {collaboration} {DELPHI}),\ }\href
  {\doibase 10.1016/S0370-2693(00)00675-4} {\bibfield  {journal} {\bibinfo
  {journal} {Phys. Lett. B}\ }\textbf {\bibinfo {volume} {485}},\ \bibinfo
  {pages} {45} (\bibinfo {year} {2000})},\ \Eprint
  {http://arxiv.org/abs/hep-ex/0103025} {arXiv:hep-ex/0103025} \BibitemShut
  {NoStop}%
\bibitem [{\citenamefont {Abdallah}\ \emph {et~al.}(2009)\citenamefont
  {Abdallah} \emph {et~al.}}]{DELPHI:2008uka}%
  \BibitemOpen
  \bibfield  {author} {\bibinfo {author} {\bibfnamefont {J.}~\bibnamefont
  {Abdallah}} \emph {et~al.} (\bibinfo {collaboration} {DELPHI}),\ }\href
  {\doibase 10.1140/epjc/s10052-009-0874-9} {\bibfield  {journal} {\bibinfo
  {journal} {Eur. Phys. J. C}\ }\textbf {\bibinfo {volume} {60}},\ \bibinfo
  {pages} {17} (\bibinfo {year} {2009})},\ \Eprint
  {http://arxiv.org/abs/0901.4486} {arXiv:0901.4486 [hep-ex]} \BibitemShut
  {NoStop}%
\bibitem [{\citenamefont {Cullen}\ and\ \citenamefont
  {Perelstein}(1999)}]{Cullen:1999hc}%
  \BibitemOpen
  \bibfield  {author} {\bibinfo {author} {\bibfnamefont {S.}~\bibnamefont
  {Cullen}}\ and\ \bibinfo {author} {\bibfnamefont {M.}~\bibnamefont
  {Perelstein}},\ }\href {\doibase 10.1103/PhysRevLett.83.268} {\bibfield
  {journal} {\bibinfo  {journal} {Phys. Rev. Lett.}\ }\textbf {\bibinfo
  {volume} {83}},\ \bibinfo {pages} {268} (\bibinfo {year} {1999})},\ \Eprint
  {http://arxiv.org/abs/hep-ph/9903422} {arXiv:hep-ph/9903422} \BibitemShut
  {NoStop}%
\bibitem [{\citenamefont {Barger}\ \emph {et~al.}(1999)\citenamefont {Barger},
  \citenamefont {Han}, \citenamefont {Kao},\ and\ \citenamefont
  {Zhang}}]{Barger:1999jf}%
  \BibitemOpen
  \bibfield  {author} {\bibinfo {author} {\bibfnamefont {V.~D.}\ \bibnamefont
  {Barger}}, \bibinfo {author} {\bibfnamefont {T.}~\bibnamefont {Han}},
  \bibinfo {author} {\bibfnamefont {C.}~\bibnamefont {Kao}}, \ and\ \bibinfo
  {author} {\bibfnamefont {R.-J.}\ \bibnamefont {Zhang}},\ }\href {\doibase
  10.1016/S0370-2693(99)00795-9} {\bibfield  {journal} {\bibinfo  {journal}
  {Phys. Lett. B}\ }\textbf {\bibinfo {volume} {461}},\ \bibinfo {pages} {34}
  (\bibinfo {year} {1999})},\ \Eprint {http://arxiv.org/abs/hep-ph/9905474}
  {arXiv:hep-ph/9905474} \BibitemShut {NoStop}%
\bibitem [{\citenamefont {Hanhart}\ \emph {et~al.}(2001)\citenamefont
  {Hanhart}, \citenamefont {Phillips}, \citenamefont {Reddy},\ and\
  \citenamefont {Savage}}]{Hanhart:2000er}%
  \BibitemOpen
  \bibfield  {author} {\bibinfo {author} {\bibfnamefont {C.}~\bibnamefont
  {Hanhart}}, \bibinfo {author} {\bibfnamefont {D.~R.}\ \bibnamefont
  {Phillips}}, \bibinfo {author} {\bibfnamefont {S.}~\bibnamefont {Reddy}}, \
  and\ \bibinfo {author} {\bibfnamefont {M.~J.}\ \bibnamefont {Savage}},\
  }\href {\doibase 10.1016/S0550-3213(00)00667-2} {\bibfield  {journal}
  {\bibinfo  {journal} {Nucl. Phys. B}\ }\textbf {\bibinfo {volume} {595}},\
  \bibinfo {pages} {335} (\bibinfo {year} {2001})},\ \Eprint
  {http://arxiv.org/abs/nucl-th/0007016} {arXiv:nucl-th/0007016} \BibitemShut
  {NoStop}%
\bibitem [{\citenamefont {Hannestad}\ and\ \citenamefont
  {Raffelt}(2001)}]{Hannestad:2001jv}%
  \BibitemOpen
  \bibfield  {author} {\bibinfo {author} {\bibfnamefont {S.}~\bibnamefont
  {Hannestad}}\ and\ \bibinfo {author} {\bibfnamefont {G.}~\bibnamefont
  {Raffelt}},\ }\href {\doibase 10.1103/PhysRevLett.87.051301} {\bibfield
  {journal} {\bibinfo  {journal} {Phys. Rev. Lett.}\ }\textbf {\bibinfo
  {volume} {87}},\ \bibinfo {pages} {051301} (\bibinfo {year} {2001})},\
  \Eprint {http://arxiv.org/abs/hep-ph/0103201} {arXiv:hep-ph/0103201}
  \BibitemShut {NoStop}%
\bibitem [{\citenamefont {Hannestad}\ and\ \citenamefont
  {Raffelt}(2003)}]{Hannestad:2003yd}%
  \BibitemOpen
  \bibfield  {author} {\bibinfo {author} {\bibfnamefont {S.}~\bibnamefont
  {Hannestad}}\ and\ \bibinfo {author} {\bibfnamefont {G.~G.}\ \bibnamefont
  {Raffelt}},\ }\href {\doibase 10.1103/PhysRevD.69.029901} {\bibfield
  {journal} {\bibinfo  {journal} {Phys. Rev. D}\ }\textbf {\bibinfo {volume}
  {67}},\ \bibinfo {pages} {125008} (\bibinfo {year} {2003})},\ \bibinfo {note}
  {[Erratum: Phys.Rev.D 69, 029901 (2004)]},\ \Eprint
  {http://arxiv.org/abs/hep-ph/0304029} {arXiv:hep-ph/0304029} \BibitemShut
  {NoStop}%
\bibitem [{\citenamefont {Hall}\ and\ \citenamefont
  {Tucker-Smith}(1999)}]{Hall:1999mk}%
  \BibitemOpen
  \bibfield  {author} {\bibinfo {author} {\bibfnamefont {L.~J.}\ \bibnamefont
  {Hall}}\ and\ \bibinfo {author} {\bibfnamefont {D.}~\bibnamefont
  {Tucker-Smith}},\ }\href {\doibase 10.1103/PhysRevD.60.085008} {\bibfield
  {journal} {\bibinfo  {journal} {Phys. Rev. D}\ }\textbf {\bibinfo {volume}
  {60}},\ \bibinfo {pages} {085008} (\bibinfo {year} {1999})},\ \Eprint
  {http://arxiv.org/abs/hep-ph/9904267} {arXiv:hep-ph/9904267} \BibitemShut
  {NoStop}%
\bibitem [{\citenamefont {Hannestad}(2001)}]{Hannestad:2001nq}%
  \BibitemOpen
  \bibfield  {author} {\bibinfo {author} {\bibfnamefont {S.}~\bibnamefont
  {Hannestad}},\ }\href {\doibase 10.1103/PhysRevD.64.023515} {\bibfield
  {journal} {\bibinfo  {journal} {Phys. Rev. D}\ }\textbf {\bibinfo {volume}
  {64}},\ \bibinfo {pages} {023515} (\bibinfo {year} {2001})},\ \Eprint
  {http://arxiv.org/abs/hep-ph/0102290} {arXiv:hep-ph/0102290} \BibitemShut
  {NoStop}%
\bibitem [{\citenamefont {Fairbairn}(2001)}]{Fairbairn:2001ct}%
  \BibitemOpen
  \bibfield  {author} {\bibinfo {author} {\bibfnamefont {M.}~\bibnamefont
  {Fairbairn}},\ }\href {\doibase 10.1016/S0370-2693(01)00501-9} {\bibfield
  {journal} {\bibinfo  {journal} {Phys. Lett. B}\ }\textbf {\bibinfo {volume}
  {508}},\ \bibinfo {pages} {335} (\bibinfo {year} {2001})},\ \Eprint
  {http://arxiv.org/abs/hep-ph/0101131} {arXiv:hep-ph/0101131} \BibitemShut
  {NoStop}%
\bibitem [{\citenamefont {Zyla}\ \emph {et~al.}(2020)\citenamefont {Zyla} \emph
  {et~al.}}]{ParticleDataGroup:2020ssz}%
  \BibitemOpen
  \bibfield  {author} {\bibinfo {author} {\bibfnamefont {P.~A.}\ \bibnamefont
  {Zyla}} \emph {et~al.} (\bibinfo {collaboration} {Particle Data Group}),\
  }\href {\doibase 10.1093/ptep/ptaa104} {\bibfield  {journal} {\bibinfo
  {journal} {PTEP}\ }\textbf {\bibinfo {volume} {2020}},\ \bibinfo {pages}
  {083C01} (\bibinfo {year} {2020})}\BibitemShut {NoStop}%
\end{thebibliography}

%merlin.mbs apsrev4-1.bst 2010-07-25 4.21a (PWD, AO, DPC) hacked
%Control: key (0)
%Control: author (72) initials jnrlst
%Control: editor formatted (1) identically to author
%Control: production of article title (-1) disabled
%Control: page (0) single
%Control: year (1) truncated
%Control: production of eprint (0) enabled
%

\end{document}